\documentclass[graybox]{svmult}

\usepackage{helvet}         
\usepackage{courier}        
\usepackage{type1cm}        
\usepackage{makeidx}         
\usepackage{graphicx}        
\usepackage{multicol}        
\usepackage[bottom]{footmisc}

\usepackage{amssymb}  
\usepackage{amsmath}
\usepackage{verbatim}
\newcommand{\dd}{\mathrm{d}}
\newcommand{\by}{by }
\newcommand{\thee}{the }
\newcommand{\as }{as }
\newcommand{\andd }{and }
\newcommand{\newc}{\newcommand}
\newc{\ra}{\rightarrow}
\newc{\lra}{\leftrightarrow}
\newc{\ov}{\overline}
\newc{\pa}{\partial}

\newc{\be}{\begin{equation}}
\newc{\ba}{\begin{eqnarray}}
\newc{\ea}{\end{eqnarray}}
\newc{\n}{\nu}

\newc{\la}{\lambda}
\newc{\e}{\epsilon}
\newc{\nn}{\nonumber}

\def\bea{\begin{eqnarray}}
\def\eea{\end{eqnarray}}
  
 \def\g{\gamma} 
 \def\d{\delta} 

\def\l{\lambda}  \def\la{\lambda} \def\m{\mu}
\def\n{\nu}

 \def\one{\mbox{1 \kern-.59em {\rm l}}}

\begin{document}

\title*{Non-commutativity in Unified Theories and Gravity$^\dagger$}
\author{G. Manolakos, G. Zoupanos}
\institute{G. Manolakos \at Physics Dept, Nat. Technical University, 157 80 Zografou, Athens, Greece,\\ \email{gmanol@central.ntua.gr}
\and G. Zoupanos \at - Physics Dept, Nat. Technical University, 157 80 Zografou, Athens, Greece \\ 
- Max-Planck Institut f\"ur Physik, F\"ohringer Ring 6,       D-80805 M\"unchen, Germany, \\ \email{George.Zoupanos@cern.ch}\\ \\ $^\dagger$ To be published in Springer Proc. Math.Stat.}
\maketitle

\abstract{First, we briefly review the Coset Space Dimensional Reduction scheme and the results of the best model so far. Then, we present the introduction of fuzzy coset spaces used as extra dimensions and perform a dimensional reduction. In turn, we describe a construction which mimics the results of a reduction, starting from a 4-dimensional theory and we present a successful example of a dynamical generation of fuzzy spheres. Finally, we propose a construction of the 3-d gravity as a gauge theory on specific non-commutative spaces.}

\section{Introduction}
\label{sec:1}

During \thee last decades, the unification of all \thee fundamental interactions has attracted \thee interest of theoretical physicists. The aim of unification led to a number of approaches and among them those that elaborate the notion of extra dimensions are particularly appealing. A consistent framework employing the idea of extra dimensions is superstring theories \cite{green-scwarz-witten} with \thee Heterotic String \cite{gross-harvey} (defined in ten dimensions) being the most promising, due to the possibility that in principle could lead to experimentally testable predictions. More specifically, the compactification of \thee $10-$dimensional spacetime \andd \thee dimensional reduction of \thee $E_8\times E_8$ initial gauge theory lead to phenomenologically interesting Grand Unified theories (GUTs), containing \thee SM gauge group.

A few years before \thee development of \thee superstring theories, another important framework with similar aims was employed, namely \thee dimensional reduction of higher-dimensional gauge theories. Pioneers in this field were Forgacs-Manton and Scherk-Schwartz studying \thee Coset Space Dimensional Reduction (CSDR) \cite{forgacs-manton,kapetanakis-zoupanos, kubyshin-mourao} \andd the group manifold reduction \cite{scherk-schwarz}, respectively. 

In both approaches, the higher-dimensional gauge fields are unifying the gauge and scalar fields, while the $4-$dimensional theory contains the surviving components after the procedure of the dimensional reduction.  Moreover, in \thee CSDR scheme, the inclusion of fermionic fields in the initial theory leads to Yukawa couplings in the $4-$dimensional theory. Furthermore, upgrading the higher-dimensional gauge theory to $\mathcal{N}=1$ supersymmetric, i.e. grouping the gauge and fermionic fields of the theory into the same vector supermultiplet, is a way to unify further the fields of the initial theory, in certain dimensions. A very remarkable achievement of the CSDR scheme is the possibility of obtaining chiral theories in four dimensions \cite{mantonns,Chapline-Slansky}.

The above context of \thee CSDR adopted some very welcome suggestions coming from \thee superstring theories (specifically from \thee Heterotic String \cite{gross-harvey}), that is \thee dimensions of the space-time and \thee gauge group of \thee higher-dimensional supersymmetric theory. In addition, taking into account \thee fact that \thee superstring theories are consistent only in ten dimensions, \thee following important issues have to be addressed, (a) distinguish \thee extra dimensions from \thee four observable ones \by considering an appropriate compactification of \thee metric and (b) determine \thee resulting $4-$dimensional theory. In addition, a suitable choice of \thee compactification manifolds could result into $\mathcal{N}=1$ supersymmetric theories in four dimensions, aiming for a chance to lead to realistic GUTs.

Requiring \thee preservation of $\mathcal{N}=1$ supersymmetry after \thee dimensional reduction, Calabi-Yau (CY) spaces serve as suitable compact, internal manifolds \cite{Candelas}. However, \thee emergence of \thee moduli stabilization problem, led to \thee study of flux compactification, in \thee context of which a wider class of internal spaces, called manifolds with $SU(3)-$structure, was suggested. In this class of manifolds, a non-vanishing, globally defined spinor is admitted. This spinor is covariantly constant with respect to a connection with torsion, versus \thee CY case, where \thee spinor is constant with respect to \thee Levi-Civita connection. Here, we consider \thee nearly-K{\"a}hler manifolds, that is an interesting class of $SU(3)-$structure manifolds \cite{cardoso-curio, Irges-Zoupanos, i-z, Butruille}. The class of homogeneous nearly-K{\"a}hler manifolds in six dimensions consists of the non-symmetric coset spaces $G_2/SU(3)$,
$Sp(4)/(SU(2)\times U(1))_{non-max}$, $SU(3)/U(1)\times U(1)$ and
the group manifold $SU(2)\times SU(2)$ \cite{Butruille} (see also
\cite{ cardoso-curio, Irges-Zoupanos, i-z}). It is worth mentioning that $4-$dimensional theories which are obtained after \thee
dimensional reduction of a $10-$dimensional $\mathcal{N}=1$
supersymmetric gauge theory over non-symmetric coset spaces, contain
supersymmetry breaking terms \cite{manousselis-zoupanos, manousselis-zoupanos2}, contrary to CY spaces.

Another very interesting framework which seems to be a natural arena for the description of physics at \thee Planck scale is the non-commutative geometry \cite{connes} - \cite{Gavriil:2015lka}. Regularizing quantum field theories, or even better, building finite ones are \thee features that render this approach as a promising framework. On \thee other hand, the construction of quantum field theories on non-commutative spaces is a difficult task and, furthermore, problematic ultraviolet features have emerged \cite{filk} (see also \cite{grosse-wulkenhaar}
and \cite{grosse-steinacker}). However, non-commutative geometry is an appropriate framework to accommodate particle models with non-commutative gauge theories \cite{connes-lott} (see also \cite{martin-bondia, dubois-madore-kerner, madorejz}).

It is remarkable that \thee two frameworks (superstring theories \andd non-commutative geometry) found contact, after \thee realization that, in M-theory and open String theory, \thee effective physics on D-branes can be described \by a non-commutative gauge theory \cite{connes-douglas-schwarz, seiberg-witten}, if a non-vanishing background antisymmetric field is present. Moreover, \thee type IIB superstring theory (\andd others related with type IIB with certain dualities) in its conjectured non-perturbative formulation as a matrix model \cite{ishibasi-kawai}, is a non-commutative theory. In \thee framework of non-commutative geometry, of particular importance is the contribution of Seiberg and Witten \cite{seiberg-witten}, which is a map between commutative \andd non-commutative gauge theories and has been the basis on which notable developments \cite{ jurco, chaichian} were achieved, including the construction of a non-commutative version of the SM  \cite{camlet}. Unfortunately, such extensions fail to solve \thee main problem of \thee SM, which is \thee presence of many free parameters.

A very interesting development in \thee framework of \thee non-commutative geometry is \thee programme in which \thee extra dimensions of higher - dimensional theories are considered to be non-commutative (fuzzy) \cite{aschieri-madore-manousselis-zoupanos} - \cite{Gavriil:2015lka}. This programme overcomes \thee ultraviolet/infrared problematic behaviours of theories defined in non-commutative spaces. A very welcome feature of such theories is that they are renormalizable, versus all known higher-dimensional theories. This aspect of \thee theory was examined from \thee $4-$dimensional point of view too, using spontaneous symmetry breakings which mimic \thee results of \thee dimensional reduction of a higher-dimensional gauge theory with non-commutative (fuzzy) extra dimensions. In addition, another interesting feature is that in theories constructed in this programme, there is an option of choosing \thee initial higher-dimensional gauge theory to be abelian. Then, non-abelian gauge theories result in lower dimensions in \thee process of \thee dimensional reduction over fuzzy coset spaces. Finally, \thee important problem of chirality in this framework has been successfully attacked \by applying an orbifold projection on an $\mathcal{N}=4$ SYM theory. After \thee orbifolding, \thee resulting theory is an $\mathcal{N}=1$ supersymmetric, chiral $SU(3)^3$.

Another interesting aspect is the study of gravity as a gauge theory on non-commutative spaces. The first and strong motivation came from Witten's work \cite{Witten:1988hc}, that (classical) $3-$d gravity with or without cosmological constant can be described as a (renormalizable) gauge theory of the isometry group of dS/AdS or Minkowski spacetime, respectively. Having already the know-how from previous works mentioned above, namely the construction of gauge theories on non-commutative spaces as extra dimensions, motivated us to study $3-$d gravity as a gauge theory on non-commutative spaces. At first, one has to determine suitable manifolds and then gauge their isometry groups, resulting with the transformations of the gauge fields and the curvature tensors. Then, one should propose an action and eventually, end up with the equations of motion \cite{ncgrav}. Our long-term goal is to obtain a $4-$d theory of gravity, hopefully with improved ultraviolet properties.  

\section{Reduction of a D-dimensional theory}
\label{sec:2}
An obvious and naive way to dimensionally reduce a higher-dimensional gauge theory is to consider all fields of the theory to be independent of the extra coordinates (trivial reduction), meaning that the Lagrangian will be independent as well. In contrast, a much more elegant way is the consideration of non-trivial dependence, that is a symmetry transformation on the fields by an element that belongs to the isometry group, $S$, of the compact coset space, $B=S/R$, formed by the extra dimensions will be a gauge transformation (symmetric fields). Therefore, the axiomatic consideration of gauge invariance of the Lagrangian, renders it independent of the extra dimensions. The above method of getting rid of the extra dimensions consists the basic concept of the CSDR scheme \cite{forgacs-manton,
kapetanakis-zoupanos, kubyshin-mourao}.

\subsection{CSDR of a D-dimensional theory}

We consider the action of a $D-$dimensional YM theory of gauge group $G$, coupled to fermions defined on $M^D$ with metric $g^{MN}$:
\begin{eqnarray}
A&=&\int
d^4xd^dy\sqrt{-g}\left[-\frac{1}{4}Tr(F_{MN}F_{K\Lambda})g^{MK}g^{N\Lambda}+\frac{i}{2}\bar
\psi \Gamma^MD_{M}\psi\right]\,,\label{actioncsdr}
\end{eqnarray}
where $D_M=\pa_M-\theta_M-A_M$, with $\theta_M=\dfrac{1}{2}\theta_{MN\Lambda}\Sigma^{N\Lambda}$ the spin connection of $M^D$ and $F_{MN}=\pa_M A_N-\pa_NA_M-[A_M,A_N]$, where $M, N, K, \Lambda=1\ldots D$ and $A_M$ and $\psi$ are
$D$-dimensional symmetric fields. The fermions can be accommodated in any representation $F$ of $G$, unless an additional symmetry, e.g. supersymmetry, is involved.

Let $\xi^{\alpha}_{A}, (A=1,...,dimS$ and $\alpha =dimR+1,...,dimS$ the curved index$)$ be \thee Killing vectors generating \thee symmetries of $S/R$ \andd $W_A$,
\thee gauge transformation associated with $\xi_A$. The
following constraint equations for scalar $\phi$, vector
$A_{\alpha}$ \andd spinor $\psi$ fields on $S/R$, derive from \thee definition of \thee symmetric fields:
\begin{eqnarray}
\delta_A\phi&=&\xi^{\alpha}_{A}\partial_\alpha\phi=D(W_A)\phi,\label{ena} \\
\delta_AA_{\alpha}&=&\xi^{\beta}_{A}\partial_{\beta}A_{\alpha}+\partial_{\alpha}\xi^{\beta}_{A}A_{\beta}=\partial_{\alpha}W_A-[W_A,A_{\alpha}], \label{duo}\\
\delta_A\psi&=&\xi^{\alpha}_{A}\partial_\alpha\psi-\frac{1}{2}G_{Abc}\Sigma^{bc}\psi=D(W_A)\psi\,,
\label{tria}
\end{eqnarray}
where $W_A$ depend only on internal coordinates $y$ and $D(W_A)$
is the gauge transformation in \thee corresponding representation, in which \thee fields are assigned. Solving \thee constraints \eqref{ena}-\eqref{tria}, one is led to \cite{forgacs-manton, kapetanakis-zoupanos} \thee unconstrained $4-$dimensional fields, as well as to \thee $4-$dimensional gauge symmetry.

We proceed with the process of \thee constraints of \thee fields of \thee theory. Gauge field $A_M$ on $M_D$ splits into its components as $(A_\mu, A_\alpha)$ corresponding to $M^4$ \andd $S/R$, respectively. Solving \thee corresponding constraint, \eqref{duo}, we get informed as follows: The $4-$dimensional gauge field, $A_\mu$, does not depend on \thee coset space coordinates and \thee $4-$dimensional gauge fields commute with \thee generators of \thee subgroup $R\in G$. This means that \thee remaining gauge symmetry, $H$, is \thee subgroup of $G$ that commutes with $R$, that is \thee centralizer of $R$ in $G$, i.e. $H=C_G(R_G)$. $A_\alpha(x,y)\equiv\phi_\alpha(x,y)$, transform as scalars in \thee $4-$dimensional theory \andd $\phi_\alpha(x,y)$ act as interwining operators connecting induced representations of $R$ acting on $G$ and $S/R$. In order to find \thee representation of scalars in \thee $4-$dimensional theory, one must decompose $G$ according to \thee following embedding:
\begin{equation}
G\supset R_G\times H\,,\qquad
adjG=(adjR,1)+(1,adjH)+\sum(r_i,h_i)\,,
\end{equation}
and $S$ under $R$:
\begin{equation}
S\supset R\,,\qquad
adjS=adjR+\sum s_i\,.
\end{equation}
We conclude that for every pair $r_i, s_i$, where $r_i$ \andd $s_i$ are identical irreducible representations of $R$, there exists a remaining scalar (Higgs) multiplet, transforming under \thee representation $h_i$ of $H$. The rest of the scalars vanish.

As for \thee spinors \cite{kapetanakis-zoupanos, mantonns,Chapline-Slansky,
Wetterich-Palla}, \thee analysis of \thee corresponding constraint, \eqref{tria}, is similar. Solving \thee constraint, one finds that spinors in \thee $4-$dimensional theory do not depend on \thee coset coordinates and act as interwining operators connecting induced  representations of $R$ in $SO(d)$ and in $G$. To obtain \thee representation under $H$ of the fermions in \thee $4-$dimensional theory, one has to decompose \thee initial representation $F$ of $G$ under \thee $R_G\times H$:
\begin{equation}
 G\supset R_G\times H\,,\qquad F=\sum (r_i,h_i),
\end{equation}
\andd \thee spinor of $SO(d)$ under $R$:
\begin{equation}
SO(d)\supset R\,,\qquad \sigma_d=\sum \sigma _j\,.
\end{equation}
Concluding, for each pair $r_i$ and $\sigma_i$, with $r_i$ and $\sigma_i$ being identical irreducible representations, there exists a multiplet, $h_i$ of spinor fields in \thee $4-$dimensional theory. If one considers Dirac fermions in \thee higher-dimensional theory, it is impossible to result with chiral fermions in four dimensions. But, if one imposes \thee  Weyl condition in \thee chiral representations of an even (in an odd higher-dimensional theory Weyl condition cannot be applied) higher-dimensional theory, eventually, one is led to a chiral $4-$dimensional theory. The most interesting case is  \thee  $D=2n+2$ higher dimensional theory, in which fermions are in \thee  adjoint representation and \thee  Weyl condition leads to two sets of chiral fermions with  \thee  same quantum numbers under $H$ of  \thee  $4-$dimensional theory. Imposing the Majorana condition, the doubling of the fermionic spectrum is lifted. In the case of $D=4n+2$ -the case of our interest-, the two conditions are compatible.

\subsection{The $4-$dimensional effective action}

We proceed with determining \thee  $4-$dimensional effective action. The first thing to do is the compactification of \thee  space $M^D$ to $M^4\times S/R$, with $S/R$ a compact coset space. After  \thee  compactification,  \thee  metric of the $M^D$ will take the following form:
\begin{equation}
g^{MN}=\left(
         \begin{array}{cc}
           \eta^{\mu \nu} & 0 \\
           0 & -g^{ab} \\
         \end{array}
       \right)\,,\label{metric_comp}
\end{equation}
where $\eta^{\mu \nu}$ is the mostly negative metric of the $4-$dimensional Minkowski spacetime and $g^{ab}$ is  \thee  metric of  \thee  coset space. Replacing \eqref{metric_comp} into \thee action, \eqref{actioncsdr}, and taking into account \thee  constraints of  \thee  fields, we obtain:
\begin{align}
A&=C\int d^4x\left[-\frac{1}{4}F_{\mu \nu}^tF^{t\mu
\nu}+\frac{1}{2}(D_\mu \phi _\alpha)^t(D^\mu \phi
^\alpha)^t+V(\phi)+\frac{i}{2}\bar \psi
\Gamma^{\mu}D_{\mu}\psi-\frac{i}{2}\bar \psi
\Gamma^{a}D_{a}\psi\right]\,,\label{tessera}
\end{align}
where $D_\mu = \partial_{\mu}-A_\mu$ and $D_a =
\partial_{a}-\theta_a-\phi_a$, with $\theta_a =
\frac{1}{2}\theta_{abc}\Sigma^{bc}$  \thee  connection of  \thee  space and $C$ \thee  volume of  \thee  space. The potential, $V(\phi)$, is given as follows:
\begin{align}
V(\phi)=-\frac{1}{4}g^{ac}g^{bd}Tr(f^C_{ab} \phi_C -[\phi_a,
\phi_b])(f^D_{cd} \phi_D - [\phi_c, \phi_d]),
\end{align}
where, $A=1,...,dimS$ and $f$'s are  \thee  structure constants
of  \thee  Lie algebra of $S$. The constraints of \thee fields, \eqref{ena}-\eqref{duo}, dictate that scalar fields, $\phi_a$, have to satisfy \thee following equation:
\begin{align}
f^D_{ai}\phi_D-[\phi_a, \phi_i]=0\,, \label{tessera_ena}
\end{align}
where $\phi_i$ are \thee generators of $R_G$. This means that some fields will be cut, while others will survive  after \thee reduction scheme and will be identified as \thee genuine Higgs fields. The potential, $V(\phi)$, expressed in terms of \thee scalars that passed the filter of the constraints (\thee  Higgs fields), is a quartic polynomial, invariant under  \thee  $4-$dimensional gauge group, $H$. Then, one has to determine \thee  vacuum (minimum of  \thee  potential) and  find out \thee  remaining gauge symmetry \cite{Chapline-Manton,Harnad,Farakos-Koutsoumbas}. In general, this is a tough procedure. However, there is a specific case in which one may result with \thee remaining symmetry, after \thee spontaneous symmetry breaking of $H$, very easily, in case the following criterion is satisfied. Whenever $S$ has an isomorphic image $S_G$ in $G$, then  \thee  $4-$dimenisonal gauge group $H$ breaks spontaneously to a subgroup $K$, where $K$ is  \thee  centralizer of $S_G$ in  \thee  gauge group of  \thee  initial, higher-dimensional, theory, $G$ \cite{kapetanakis-zoupanos,Chapline-Manton,Harnad,Farakos-Koutsoumbas}. This is demonstrated in the following scheme:
\begin{align}
  G\supset &\,S_G\times K\nonumber\\
  &\,\cup\quad\,\,\,\cap    \nonumber\\
  G\supset&\,R_G\times H
\end{align}
The potential of  \thee  resulting $4-$dimensional gauge theory is always of spontaneous symmetry breaking form, when  \thee  coset space is symmetric\footnote{A coset space is called symmetric when $f_{ab}^c=0$}. It is rather unpleasant, that in this case, after  \thee  application of the reduction scheme, \thee  fermions obtained are supermassive -as in  \thee  Kaluza-Klein theory-. 

Let us now demonstrate some results coming from  \thee
dimensional reduction of  \thee  $\mathcal{N}=1, E_8$ SYM over  \thee  nearly-K{\"a}hler manifold $SU(3)/U(1)\times U(1)$. The $4-$dimensional gauge group is obtained \by decomposing $E_8$ under $R=U(1)\times U(1)$, as follows:
\begin{equation}
E_8\supset E_6\times SU(3)\supset E_6\times U(1)_A\times U(1)_B\,.
\end{equation}
Satisfying \thee criterion mentioned above, \thee resulting $4-$dimensional gauge group is:
\begin{equation}
H=C_{E_8}(U(1)_A\times U(1)_B)=E_6\times U(1)_A\times U(1)_B\,.
\end{equation}
The decomposition of \thee adjoint representation of $E_8$, the $248$, under $U(1)_A\times U(1)_B$ gives the surviving scalar \andd fermion fields. After the application of \thee CSDR rules, one obtains  \thee  resulting $4$-dimensional theory, which is an $\mathcal{N}=1$, $E_6$ GUT, with $U(1)_A, U(1)_B$ global symmetries. The potential is determined after a lengthy calculation found in ref \cite{manousselis-zoupanos2}. Apart from \thee  $F-$ and $D-$ terms contributing to this potential, one can determine also scalar masses and trilinear scalar terms, identified with  \thee  scalar part of  \thee  soft supersymmetry breaking sector of  \thee  theory. In addition,  \thee  gaugino becomes massive, receiving a contribution from  \thee  torsion, unlike  \thee  rest soft supersymmetry breaking terms. It is worth-noting that \thee  CSDR scheme leads straight to \thee  soft supersymmetry breaking sector without any additional assumption.

Further breaking of \thee  $E_6$ GUT is achieved by \thee  Wilson flux mechanism. Details for the present case can be found in ref \cite{i-z}. The theory derived is a softly broken $\mathcal{N}=1$, chiral $SU(3)^3$ theory which can be further broken to an extension of  \thee  MSSM.

\section{Fuzzy spaces}

A particularly interesting framework, in which particle (and gravity) models can be built on, is non-commutative geometry. For now, we focus on the fuzzy spaces (non-commutative spaces defined as matrix approximations of continuous manifolds), which will be used as extra dimensions in a higher dimensional theory. In this section we give details about the definition of a specific fuzzy space, the fuzzy sphere and the differential calculus on it. Then, we briefly present how to do gauge theory on this matrix-approximated sphere, concluding all the necessary information for the applications of the next section.   
\subsection{The Fuzzy sphere}
 We will introduce \thee fuzzy sphere, $S^2_N$ \cite{Madore:1991bw}, through a modification of \thee familiar, ordinary sphere $S^2$, which is considered as a manifold embedded into  \thee  $3-$dimensional Euclidean space, $\mathbb{R}^3$. This embedding allows the specification of \thee  algebra of functions on $S^2$ through $\mathbb{R}^3$, \by imposing  \thee  constraint
\begin{equation}
  \sum_{a=1}^3x_a^2=R^2\,,\label{spherecondition}
\end{equation}
where $x_a$ are  \thee  coordinates of $\mathbb{R}^3$ and $R$  \thee  radius of the sphere. The isometry group of $S^2$ is a global $SO(3)$, generated \by \thee three angular momentum operators, $L_a=-i\epsilon_{abc}x_b\pa_c$, due to  \thee  isomorphism $SO(3)\simeq SU(2)$. Writing \thee three generators, $L_a$, in terms of  \thee  spherical coordinates $\theta,\phi$, they are expressed as $L_a=-i\xi_a^\alpha\pa_\alpha$,
where \thee greek index, $\alpha$, denotes \thee spherical coordinates and $\xi_a^\alpha$ are \thee components of \thee Killing vector fields, which generate \thee isometries of \thee sphere\footnote{The $S^2$ metric can be expressed in terms of \thee Killing vectors as $g^{\alpha\beta}=\dfrac{1}{R^2}\xi^\alpha_a\xi_a^\beta$.}.

The operator defined as:
\begin{equation}
  L^2=-R^2\triangle_{S^2}=-R^2\frac{1}{\sqrt{g}}\pa_a(g^{ab}\sqrt{g}\pa_b)\,,
\end{equation}
has the spherical harmonics, $Y_{lm}(\theta,\phi)$ as eigenfunctions. In order to calculate the eigenvalues of $L^2$, one has to act on $Y_{lm}(\theta,\phi)$:
\begin{equation}
  L^2 Y_{lm}=l(l+1)Y_{lm}\,,
\end{equation}
with $l$ being a positive integer. The eigenfunctions $Y_{lm}(\theta,\phi)$ satisfy \thee orthogonality condition:
\begin{equation}
  \int \sin\theta d\theta d\phi Y_{lm}^\dag
  Y_{l'm'}=\delta_{ll'}\delta_{mm'}\,.
\end{equation}
$Y_{lm}(\theta,\phi)$ form a complete and orthogonal set of functions, therefore any function on $S^2$ can be expanded on them:
\begin{equation}
  a(\theta,\phi)=\sum_{l=0}^\infty\sum_{m=-l}^la_{lm}Y_{lm}(\theta,\phi)\,,\label{sphereexpansion}
\end{equation}
where $a_{lm}$ are complex coefficients. In an alternative way, spherical harmonics can be expressed in terms of \thee coordinates $x_a$, as:
\begin{equation}
  Y_{lm}(\theta,\phi)=\sum_{\vec{a}}f^{lm}_{a_1\ldots a_l}x^{a_1\ldots
  a_l}\,,\label{cartesianharmonics}
\end{equation}
where $f^{lm}_{a_1\ldots a_l}$ is an $l-$rank (traceless) symmetric tensor.

Let us now modify the above, in order to obtain the fuzzy version of $S^2$. Fuzzy sphere is a typical case of a non-commutative space, meaning that functions do not commute, contrary to the $S^2$ case, with $l$ having an upper limit. Therefore, this truncation yields a finite dimensional (non-commutative) algebra, $l^2$ dimensional. Thus, it is natural to consider \thee truncated algebra as a matrix algebra and it is consistent to consider \thee fuzzy sphere as a matrix approximation of \thee $S^2$. According to \thee above, $N$-dimensional matrices are expanded on a fuzzy sphere as:
\begin{equation}
  \hat{a}=\sum_{l=0}^{N-1}\sum_{m=-l}^{l}a_{lm}\hat{Y}_{lm}\,,\label{fuzzyexpansion}
\end{equation}
where $\hat{Y}_{lm}$ are spherical harmonics of \thee fuzzy sphere, given \by:
\begin{equation}
  \hat{Y}_{lm}=R^{-l}\sum_{\vec{a}}f_{a_1\ldots
  a_l}^{lm}\hat{X}^{a_1}\cdots\hat{X}^{a_l}\,,
\end{equation}
where:
\begin{equation}
  \hat{X}_a=\frac{2R}{\sqrt{N^2-1}}\lambda_a^{(N)}\,,\label{fuzzycoordinates}
\end{equation}
where $\lambda_a^{(N)}$ are \thee $SU(2)$ generators in \thee
$N$-dimensional representation and $f_{a_1\ldots a_l}^{lm}$ is \thee same tensor, used in \eqref{cartesianharmonics}. The $\hat{Y}_{lm}$ also satisfy \thee orthonormality condition:
\begin{equation}
  \text{Tr}_N\left(\hat{Y}_{lm}^\dag\hat{Y}_{l'm'}\right)=\delta_{ll'}\delta_{mm'}\,.
\end{equation}
Moreover, there is a correspondence between \thee expansion of a function, \eqref{sphereexpansion}, and that of a matrix,
\eqref{fuzzyexpansion}, on \thee ordinary and \thee fuzzy sphere,
respectively:
\begin{equation}
  \hat{a}=\sum_{l=0}^{N-1}\sum_{m=-l}^{l}a_{lm}\hat{Y}_{lm}\quad\rightarrow\quad
  a=\sum_{l=0}^{N-1}\sum_{m=-l}^{l}a_{lm}Y_{lm}(\theta,\phi)\,.\label{mapping}
\end{equation}
The above obviously maps matrices to functions. The introduction of \thee fuzzy sphere as a truncation of \thee algebra
of functions on $S^2$, suggests, as a natural choice (but not unique), the consideration of \thee same $a_{lm}$. The above is a $1:1$ mapping given \by ref \cite{andrews}:
\begin{equation}
  a(\theta, \phi)=\sum_{lm}\text{Tr}_N(\hat{Y}_{lm}^\dag\hat{a})Y_{lm}(\theta, \phi)\,,
\end{equation}
while the matrix trace is mapped to an integral over \thee sphere:
\begin{equation}
  \frac{1}{N}\text{Tr}_N\quad\rightarrow\quad\frac{1}{4\pi}\int d\Omega\,.
\end{equation}
To sum up, the fuzzy sphere is a matrix approximation of $S^2$. The price one has to pay for the truncation of \thee algebra of functions is the loss of commutativity, yielding the non-commutative algebra of matrices, $\text{Mat}(N;\mathbb{C})$. Therefore, \thee fuzzy sphere, $S_N^2$, is \thee non-commutative manifold with $\hat{X}_a$ being \thee coordinate functions. As given \by \eqref{fuzzycoordinates}, $\hat{X}_a$ are $N\times N$ hermitian matrices produced \by \thee generators of $SU(2)$ in \thee $N-$dimensional representation. It is obvious that they have to obey both \thee condition:
\begin{equation}
\sum_{a=1}^3\hat{X}_a\hat{X}_a=R^2\,,\label{fuzzyspherecondition}
\end{equation}
which is \thee equivalent of \eqref{spherecondition} and \thee
commutation relation:
\begin{equation}
[\hat{X}_a,\hat{X}_b]=i\alpha\epsilon_{abc}\hat{X}_c\,,\quad
  \alpha=\frac{2R}{\sqrt{N^2-1}}\,.\label{fuzzycommutationrelation}
\end{equation}
It is equivalent to consider \thee description of the algebra on $S_N$ \by \thee antihermitian matrices:
\begin{equation}
  X_a=\frac{\hat{X}_a}{i\alpha R}\,,
\end{equation}
also satisfying a variation of \eqref{fuzzyspherecondition},
\eqref{fuzzycommutationrelation}:
\begin{equation}
  \sum_{a=1}^3X_aX_a=-\frac{1}{\alpha^2}\,,\quad
  [X_a,X_b]=C_{abc}X_c\,,
\end{equation}
where $C_{abc}=\dfrac{1}{R}\epsilon_{abc}$\,.

Let us proceed \by giving a short description of \thee differential calculus on \thee fuzzy sphere, which is  $3-$dimensional and $SU(2)$ covariant. The derivations of a function $f$, along $X_a$ are:
\begin{equation}
  e_a(f)=[X_a,f]\,,
\end{equation}
and, consequently, \thee Lie derivative on $f$ is:
\begin{equation}
  \mathcal{L}_af=[X_a,f]\,,\label{liederonfunction}
\end{equation}
where $\mathcal{L}_a$ obeys both \thee Leibniz rule and \thee commutation relation of the $SU(2)$ algebra:
\begin{equation}
  [\mathcal{L}_a,\mathcal{L}_b]=C_{abc}\mathcal{L}_c\,.\label{liecommutator}
\end{equation}
Working with differential forms, let $\theta^a$ be \thee $1-$forms dual to \thee vector fields $e_a$, namely $\langle e_a,\theta^b\rangle=\delta^b_a$. Therefore, \thee action of the exterior derivative, $d$ on a function $f$, gives:
\begin{equation}
  df=[X_a,f]\theta^a\,,
\end{equation}
while \thee action of \thee Lie derivative on \thee $1-$forms
$\theta^b$ gives:
\begin{equation}
  \mathcal{L}_a\theta^b=C_{abc}\theta^c\,.\label{liederonform}
\end{equation}
The Lie derivative obeys \thee Leibniz rule, therefore action on any $1-$form $\omega=\omega_a\theta^a$ gives:
\begin{equation}
  \mathcal{L}_b\omega=\mathcal{L}_b(\omega_a\theta^a)=[X_b,\omega_a]\theta^a-\omega_aC^a_{bc}\theta^c\,,
\end{equation}
where we have applied \eqref{liederonfunction} and
\eqref{liederonform}. Therefore, one obtains \thee result:
\begin{equation}
  (\mathcal{L}_b\omega)_a=[X_b,\omega_a]-\omega_cC^c_{ba}\,.
\end{equation}

After the description of \thee differential geometry of the fuzzy sphere, one could move on to \thee study of \thee differential geometry of $M^4\times S^2_N$, that is \thee product of Minkowski spacetime and fuzzy sphere with fuzziness level $N-1$. For example, any $1-$form $A$ of $M^4\times
S^2_N$ can be expressed in terms of $M^4$ and $S_N^2$, that
is:
\begin{equation}
  A=A_\mu dx^\mu+A_a\theta^a\,,\label{gaugepotentialoneform}
\end{equation}
where $A_\mu, A_a$ depend on $x^\mu$ and $X_a$ coordinates.

In addition, instead of functions, one may consider spinors on \thee $S_N^2$ \cite{aschieri-madore-manousselis-zoupanos}. Moreover, there are studies of \thee differential geometry of various higher-dimensional fuzzy spaces, e.g. of the fuzzy $CP^M$ \cite{aschieri-madore-manousselis-zoupanos}.

\subsection{Gauge theory on fuzzy sphere}

Let us consider a field $\phi(X_a)$ on \thee $S_N^2$, depending on \thee powers of $X_a$ \cite{madore-schrami-schup-wess}. The infinitesimal transformation of $\phi(X_a)$ is given by: 
\begin{equation}
  \delta\phi(X)=\lambda(X)\phi(X)\,,\label{gaugetransffuzzy}
\end{equation}
where $\lambda(X)$ is \thee gauge parameter. If $\lambda(X)$ is an antihermitian function of $X_a$, \thee \eqref{gaugetransffuzzy} is an infinitesimal (abelian) $U(1)$ transformation, while if $\lambda(X)$ is valued in $\text{Lie}(U(P))$ (\thee algebra of $P\times P$ hermitian matrices), then \eqref{gaugetransffuzzy} is the infinitesimal (non-abelian), $U(P)$. Also, $\delta X_a=0$, that is a condition which ensures \thee invariance of \thee coordinates under a gauge
transformation. Therefore, the left multiplication by a coordinate is not a covariant operation:
\begin{equation}
  \delta(X_a\phi)=X_a\lambda(X)\phi\,,
\end{equation}
and in general it holds that:
\begin{equation}
  X_a\lambda(X)\phi\neq\lambda(X)X_a\phi\,.
\end{equation}
Inspired \by \thee non-fuzzy gauge theory, one may proceed with the introduction of \thee covariant coordinates, $\phi_a$, such that:
\begin{equation}
  \delta(\phi_a\phi)=\lambda\phi_a\phi\,,
\end{equation}
which holds if:
\begin{equation}
  \delta(\phi_a)=[\lambda,\phi_a]\,.\label{transformationofphia}
\end{equation}
Usual (non-fuzzy) gauge theory also suggests the definition:
\begin{equation}
  \phi_a\equiv X_a+A_a\,,\label{covariantfield}
\end{equation}
with $A_a$ being interpreted as \thee gauge potential of \thee
non-commutative theory. Therefore, \thee covariant coordinates, $\phi_a$, are \thee non-commutative analogue of \thee covariant derivative encountered in ordinary gauge theories. From \eqref{transformationofphia}, \eqref{covariantfield}, one is led to \thee gauge transformation of $A_a$:
\begin{equation}
  \delta A_a=-[X_a,\lambda]+[\lambda,A_a]\,,
\end{equation}
a form that encourages \thee interpretation of $A_a$ as a gauge field. Then, it is natural to define a field strength tensor, $F_{ab}$, as:
\begin{equation}
  F_{ab}\equiv[X_a,A_b]-[X_b,A_a]+[A_a,A_b]-C^c_{ab}A_c=[\phi_a,\phi_b]-C^c_{ab}\phi_c\,.
\end{equation}
It can be proven that \thee field strength tensor transforms covariantly:
\begin{equation}
  \delta F_{ab}=[\lambda,F_{ab}]\,.
\end{equation}
\section{Dimensional reduction of higher-dimensional theory with fuzzy extra dimensions}
In this section we present the ordinary (naive) dimensional reduction of a higher-dimensional theory with fuzzy extra dimensions and the coset space dimensional reduction, adjusted to the non-commutative framework. 

\subsection{Ordinary fuzzy dimensional reduction}
Let us now apply the structure of the previous section, considering a higher-dimensional theory with fuzzy extra dimensions and then perform a simple (trivial) dimensional reduction. The higher-dimensional theory is defined on $M^4\times (S/R)_F$, with $(S/R)_F$ a fuzzy coset, e.g. \thee fuzzy sphere, $S^2_N$, with symmetry governed by the gauge group $G=U(P)$. The Y-M action is:
\begin{equation}
  \mathcal{S}_{YM}=\frac{1}{4g^2}\int
  d^4xk\text{Tr}\text{tr}_GF_{MN}F^{MN}\,,\label{actiondimred}
\end{equation}
with $\text{tr}_G$ \thee trace over the generators of the gauge group, $G$, and $k\text{Tr}$\footnote{In general, $k$ is a parameter related to \thee
size of \thee fuzzy coset space. In \thee case of \thee fuzzy
sphere, $k$ is related to \thee radius of \thee
sphere and \thee integer $l$.} \thee integration over $(S/R)_F$,
i.e. \thee fuzzy coset described \by $N\times N$ matrices and
$F_{MN}$ \thee higher-dimensional field strength tensor, which is composed of both $4-$dimensional spacetime and extra-dimensional parts, i.e. $(F_{\mu\nu},F_{\mu a},F_{ab})$.
The fuzzy extra-dimensional components of $F_{MN}$  are expressed in terms of \thee covariant coordinates $\phi_a$:
\begin{align}
  F_{\mu a}&=\pa_\mu\phi_a+[A_\mu,\phi_a]=D_\mu\phi_a\nonumber\\
  F_{ab}&=[X_a,A_b]-[X_b,A_a]+[A_a,A_b]-C_{ba}^cA_{ac}\,.\nonumber
\end{align}
Replacing \thee above equations in \eqref{actiondimred}, \thee action becomes:
\begin{equation}
  \mathcal{S}_{YM}=\int
  d^4x\text{Tr}\text{tr}_G\left(\frac{k}{4g^2}F_{\mu\nu}^2+\frac{k}{2g^2}(D_\mu\phi_a)^2-V(\phi)\right)\,,\label{action2}
\end{equation}
where $V(\phi)$ is \thee potential, derived from \thee kinetic
term of $F_{ab}$, that is
\begin{align}
  V(\phi)&=-\frac{k}{4g^2}\text{Tr}\text{tr}_G\sum_{ab}F_{ab}F_{ab}\nonumber\\
  &=-\frac{k}{4g^2}\text{Tr}\text{tr}_G\left([\phi_a,\phi_b][\phi^a,\phi^b]-4C_{abc}\phi^a\phi^b\phi^c+2R^{-2}\phi^2\right)\,.
\end{align} 
The \eqref{action2} admits a natural interpretation as an action of a $4-$dimensional theory. Let $\lambda(x^\mu,X^a)$ be \thee gauge parameter of an infinitesimal gauge transformation of $G$. This transformation can be viewed as a $M^4$ gauge transformation:
\begin{equation}
  \lambda(x^\mu,X^a)=\lambda^I(x^\mu,X^a)\mathcal{T}^I=\lambda^{h,I}(x^\mu)\mathcal{T}^h\mathcal{T}^I\,,\label{lambdareduction}
\end{equation}
where $\mathcal{T}^I$ denote \thee hermitian generators of \thee gauge group $U(P)$ and $\lambda^I(x^\mu,X^a)$ are \thee $N\times N$ antihermitian matrices, which means that they can be expressed as $\lambda^{I,h}(x^\mu)\mathcal{T}^h$, where $\mathcal{T}^h$ are \thee antihermitian generators of $U(N)$ and $\lambda^{I,h}(x^\mu), h=1,\ldots, N^2$, are \thee Kaluza-Klein modes of $\lambda^I(x^\mu,X^a)$. In turn, we can assume that \thee fields on \thee right hand side of \eqref{lambdareduction} could be considered as a field valued in \thee tensor product Lie algebra $\text{Lie}\left(U(N)\right)\otimes\text{Lie}\left(U(P)\right)$, that is \thee algebra
$\text{Lie}\left(U(NP)\right)$. Similar consideration applies for the \thee gauge field $A_\nu$, too:
\begin{equation}
  A_\nu(x^\mu,X^a)=A_\nu^I(x^\mu,X^a)\mathcal{T}^I=A_\nu^{h,I}(x^\mu)\mathcal{T}^h\mathcal{T}^I\,,
\end{equation}
which can be regarded as a gauge field on $M^4$ taking values in \thee $\text{Lie}\left(U(NP)\right)$ algebra. A similar
consideration can be applied for the scalars, too\footnote{Also, $\text{Tr}\text{tr}_G$ is interpreted as
\thee trace of \thee $U(NP)$ matrices.}.

A very important feature of the above structure is \thee enhancement of \thee gauge symmetry of \thee $4-$dimensional theory as compared to \thee symmetry of \thee starting, higher-dimensional theory. Specifically, one may choose to start with abelian gauge group in higher dimensions and result with a non-abelian gauge symmetry in four dimensions. An undesirable result is that \thee scalars are accommodated in \thee adjoint representation of \thee $4-$dimensional gauge group, meaning that they cannot trigger \thee electroweak symmetry breaking. In order to overcome this drawback, one should try to employ an alternative dimensional reduction.

\subsection{Fuzzy CSDR}

There is an alternative way to obtain a $4-$dimensional gauge theory from a higher-dimensional theory. This is realized by performing a non-trivial dimensional reduction, which, in our case, is \thee CSDR, modified as it must, since the extra dimensions are now fuzzy coset spaces \cite{aschieri-madore-manousselis-zoupanos}\footnote{See also \cite{Harland-kurkcuoglu}.}. 

We begin by presenting the similarities of the two reduction schemes: CSDR and fuzzy CSDR. The first similarity is that fuzziness does not affect \thee isometries and the second one is that gauge couplings defined on both spaces have \thee same dimensionality.

A major difference between fuzzy and ordinary CSDR is that the $4 -$
dimensional gauge group appearing in the ordinary CSDR after the geometrical breaking and before the spontaneous symmetry breaking -due to the $4-$dimensional Higgs fields- does not appear in the fuzzy CSDR. In the latter, the spontaneous symmetry breaking takes already place by solving the fuzzy CSDR constraints and the $4-$dimensional potential appears already shifted to a minimum. Therefore, in four dimensions, appears only the physical Higgs field that survives after a spontaneous symmetry breaking. Correspondingly, in the Yukawa sector of the theory we have results of the spontaneous symmetry breaking, i.e. massive fermions and Yukawa interactions among fermions and the physical Higgs field. We conclude that if one would like to describe the spontaneous symmtery breaking of the SM in the present framework, then one would be naturally led to large extra dimensions. Another fundamental difference between the two CSDR reductions is the fact that a non-Abelian gauge group, $G$, is not required in many dimensions. Indeed, it turns out that the presence of a $U(1)$ in the higher-dimensional theory is enough to obtain non-Abelian gauge theories in four dimensions. Another fundamental difference as compared to all known theories defined in more than four dimensions is that the present ones are renormalizable. For technicalities, one should consult the original paper or some review papers \cite{aschieri-madore-manousselis-zoupanos}.

\section{Orbifolds and fuzzy extra dimensions}
The involvement of \thee orbifold structure (similar to \thee one developed in \cite{kachru-silverstein}) in \thee framework of gauge theories with fuzzy coset spaces as extra dimensions, was suggested in order to obtain chiral low-energy theories. In order to support \thee renormalizability of \thee theories constructed so far using fuzzy extra dimensions \thee reverse procedure was considered, that is to start from a renormalizable theory in four dimensions and reproduce \thee results of a higher-dimensional theory reduced over fuzzy coset spaces \cite{aschieri-grammatikopoulos, steinacker-zoupanos, chatzistavrakidis-steinacker-zoupanos}. This idea was realized as follows: one starts with a gauge theory in four dimensions with an appropriate set of scalar fields and a suitable potential, which leads to vacua that could be identified as -dynamically generated- fuzzy extra dimensions, including a finite Kaluza-Klein tower of massive modes. This reverse procedure is targeting at proving that an initial abelian gauge theory does not have to be considered in higher dimensions, with \thee non-abelian gauge theory structure emerging from fluctuations of \thee coordinates \cite{steinacker1, kim}. The whole idea share some similarities with \thee idea of dimensional deconstruction, introduced earlier \cite{ArkaniHamed:2001nc, ArkaniHamed:2001ca}.

Then, there was an attempt to include fermions, but \thee best one could achieve (for some time) contained mirror fermions in bifundamental representations of \thee low-energy gauge group \cite{steinacker-zoupanos,chatzistavrakidis-steinacker-zoupanos}. Although mirror fermions do not exclude \thee possibility to make contact with phenomenology \cite{maalampi-roos}, it is preferable to result with exactly chiral fermions.

In the following, we are going to deal with \thee $\mathbb{Z}_3$ orbifold projection of \thee $\mathcal{N}=4$ Supersymmetric Yang Mills (SYM) theory \cite{brink-schwarz-scherk}, studying \thee action of \thee discrete group on \thee fields of \thee theory and \thee emerging superpotential in \thee projected theory \cite{fuzzy}.

\subsection{$\mathcal{N}=4$ SYM field theory and $\mathbb{Z}_3$ orbifolds \label{subsec:subsection}}

Let us consider an $\mathcal{N}=4$ supersymmetric $SU(3N)$
gauge theory defined on \thee Minkowski spacetime with a particle spectrum of \thee theory (in \thee $\mathcal{N}=1$ terminology) that consists of an $SU(3N)$ gauge
supermultiplet and three adjoint chiral supermultiplets
$\Phi^i\,,i=1,2,3$. The component fields of \thee above
supermultiplets are \thee gauge bosons, $A_\mu,\,\mu=1,\ldots,4$, six adjoint real (or three complex) scalars $\phi^a,\,a=1,\ldots,6$ and four adjoint Weyl fermions $\psi^p,\,p=1,\ldots,4$. The scalars and Weyl fermions transform according to \thee $6$ and $4$ representations of \thee $SU(4)_R$ $R$-symmetry of \thee theory, respectively, while \thee gauge bosons are singlets. For the introduction of the orbifolds, \thee discrete group $\mathbb{Z}_3$ has to be considered as a subgroup of $SU(4)_R$. The $\mathbb{Z}_3$ is not embedded into $SU(4)_R$ in a unique way with \thee options not being equivalent, since \thee choice of embedding affects \thee amount of \thee remnant supersymmetry \cite{kachru-silverstein}:
\begin{itemize}
  \item Maximal embedding of $\mathbb{Z}_3$ into $SU(4)_R$ would lead to non-supersymmetric models, therefore it is excluded.
  \item Embedding $\mathbb{Z}_3$ in a subgroup of $SU(4)_R$:
  \begin{itemize}
    \item[-] Embedding into an $SU(2)$ subgroup would lead to
    $\mathcal{N}=2$ supersymmetric models with $SU(2)_R$ $R$-symmetry.
    \item[-] Embedding into an $SU(3)$ subgroup would lead to
    $\mathcal{N}=1$ supersymmetric models with $U(1)_R$
    $R$-symmetry.
  \end{itemize}
\end{itemize}

We focus on \thee last embedding, which gives the desired remnant supersymmetry. Let us consider a generator $g\in\mathbb{Z}_3$, labeled (for convenience) \by three
integers $\vec{a}=(a_1, a_2, a_3)$ \cite{douglas-greene-morrison} satisfying \thee relation
\begin{equation}
  a_1+a_2+a_3=0\,\,\text{mod}\,3\,.
\end{equation}
The last equation implies that $\mathbb{Z}_3$ is embedded
in \thee $SU(3)$ subgroup, i.e. \thee remnant supersymmetry is \thee desired $\mathcal{N}=1$ \cite{bailin-love}. Since \thee various fields of \thee theory transform differently under $SU(4)_R$, $\mathbb{Z}_3$ will act
non-trivially on them. Gauge and gaugino fields are singlets under $SU(4)_R$, therefore \thee geometric action of \thee $\mathbb{Z}_3$ rotation is trivial. The action of $\mathbb{Z}_3$ on \thee complex scalar fields is given \by \thee matrix $\gamma(g)_{ij}=\delta_{ij}\omega^{a_i}$, where $\omega=e^{\frac{2\pi}{3}}$ and \thee action of $\mathbb{Z}_3$ on \thee fermions $\phi^i$ is given \by $\gamma(g)_{ij}=\delta_{ij}\omega^{b_i}$, where $b_i=-\dfrac{1}{2}(a_{i+1}+a_{i+2}-a_i)$\footnote{Also modulo 3}. In \thee present case, \thee three integers of \thee generator $g$ are $(1,1,-2)$, meaning that $a_i=b_i$. The matter fields are not gauge invariant, therefore $\mathbb{Z}_3$ acts on their gauge indices, too. The action of this rotation is given \by \thee matrix
\begin{equation}
  \gamma_3=\left(
             \begin{array}{ccc}
               \mathbf{1}_N & 0 & 0 \\
               0 & \omega\mathbf{1}_N & 0 \\
               0 & 0 & \omega^2\mathbf{1}_N \\
             \end{array}
           \right)\,.\label{gammatria}
\end{equation}
There is no specific reason for these blocks to have \thee same
dimensionality (see e.g.\cite{aldabaz-ibanez,
lawrence-nekrason-vafa, kiritsis}). However, it is the same, because \thee projected theory must be free of anomalies.

After \thee orbifold projection, \thee spectrum of \thee theory consists of \thee fields that are invariant under \thee combined action of \thee discrete group, $\mathbb{Z}_3$, on \thee ``geometric''\footnote{In case of ordinary reduction of a $10$-dimensional $\mathcal{N}=1$ SYM theory, one obtains an $\mathcal{N}=4$ SYM Yang-Mills theory in four dimensions having a global $SU(4)_R$ symmetry which is identified with \thee tangent space $SO(6)$ of \thee extra dimensions \cite{manousselis-zoupanos, manousselis-zoupanos2, Sohnius:1985qm}.} and gauge indices
\cite{douglas-greene-morrison}. As for \thee gauge bosons, \thee projection is $A_\mu=\gamma_3 A_\mu\gamma_3^{-1}$.
Therefore, taking into consideration \eqref{gammatria}, \thee gauge group of \thee initial theory breaks down to \thee group $H=SU(N)\times SU(N)\times SU(N)$ in \thee projected theory.
The complex scalar fields transform non-trivially under \thee gauge and $R-$symmetry, so \thee
projection is $\phi_{IJ}^i=\omega^{I-J+a_i}\phi^i_{IJ}$,
where $I,J$ are gauge indices. Therefore, $J=I+a_i$, meaning
that \thee scalar fields surviving \thee orbifold projection have \thee form $\phi_{I,J+a_i}$ and transform under \thee gauge group, $H$, as:
\begin{equation}
  3\cdot\left((N,\bar{N},1)+(\bar{N},1,N)+(1,N,\bar{N})\right)\,.\label{repaftrprojection}
\end{equation}
Similarly, fermions transform non-trivially under both gauge group and $R-$symmetry, with \thee projection being $\psi^i_{IJ}=\omega^{I-J+b_i}\psi_{IJ}^i$. Therefore, \thee fermions surviving \thee projection have \thee form $\psi^i_{I,I+b_i}$ accommodated in \thee same representation as \thee scalars, \eqref{repaftrprojection}, demonstrating \thee $\mathcal{N}=1$ remnant supersymmetry. It is notable that \thee representations \eqref{repaftrprojection} of \thee
resulting theory are anomaly free.  So, in a nutshell, fermions are accommodated into chiral representations of $H$, divided into three generations since \thee particle spectrum contains three $\mathcal{N}=1$ chiral supermultiplets.

The interactions of \thee projected model are included in the superpotential. To specify it, one has to start with \thee superpotential of \thee $\mathcal{N}=4$ SYM theory \cite{brink-schwarz-scherk}:
\begin{equation}
  W_{\mathcal{N}=4}=\epsilon_{ijk}\text{Tr}(\Phi^i\Phi^j\Phi^k)\,,
\end{equation}
where, $\Phi^i,\Phi^j,\Phi^k$ are \thee three chiral superfields
of \thee theory. After \thee projection, \thee structure of \thee superpotential remains the same, encrypting only \thee interactions among \thee surviving fields of \thee $\mathcal{N}=1$ theory:
\begin{equation}
  W_{\mathcal{N}=1}^{(proj)}=\sum_I\epsilon_{ijk}\Phi_{I,I+a_i}^i\Phi_{I+a_i,I+a_i+a_j}^j\Phi_{I+a_i+a_j,I}^k\,.\label{projectedsuper}
\end{equation}

\subsection{Dynamical generation of twisted fuzzy
spheres}

The superpotential $W_{\mathcal{N}=1}^{proj}$, \eqref{projectedsuper}, produces \thee scalar potential:
\begin{equation}
  V_{\mathcal{N}=1}^{proj}(\phi)=\frac{1}{4}\text{Tr}\left([\phi^i,\phi^j]^\dag[\phi^i,\phi^j]\right)\,,\label{scalarpotential}
\end{equation}
where, $\phi^i$ are \thee scalar components of \thee
superfield, $\Phi^i$. The potential $V_{\mathcal{N}=1}^{proj}(\phi)$ gets minimized \by vanishing vevs of \thee fields, so, in order to result with solutions interpreted as vacua of a non-commutative geometry, some modifications have to take place. So, in order to result with minima of $V_{\mathcal{N}=1}^{proj}(\phi)$, 
soft $\mathcal{N}=1$ supersymmetric terms of \thee form\footnote{The
SSB terms that will be inserted into $V_{\mathcal{N}=1}^{proj}(\phi)$, are purely scalar. Although this is enough for our purpose, it is obvious that more SSB terms have to be included too, in order to obtain \thee full SSB sector \cite{djouadi}.}
\begin{equation}
  V_{SSB}=\frac{1}{2}\sum_im_i^2\phi^{i\dag}\phi^i+\frac{1}{2}\sum_{i,j,k}h_{ijk}\phi^i\phi^j\phi^k+h.c.\label{ssbterms}
\end{equation}
are included, with $h_{ijk}=0$ unless
$i+j+k\equiv0\,\text{mod}3$. The introduction of SSB terms does not cause embarassement, since \thee presence of an SSB sector is necessary anyway for a model to have phenomenological viability, see e.g.\cite{djouadi}. The $D$-terms of \thee theory are given \by
\begin{equation}
  V_D=\frac{1}{2}D^2=\frac{1}{2}D^ID_I\,,
\end{equation}
where $D^I=\phi_i^\dag T^I\phi^i$, where $T^I$ are \thee generators, accommodated in \thee representation of \thee corresponding chiral multiplets. Therefore, putting all potential terms together, \thee total potential of \thee theory is:
\begin{equation}
V=V_{\mathcal{N}=1}^{proj}+V_{SSB}+V_D\,.\label{totalscalarpotential}
\end{equation}
An appropriate choice of \thee parameters $m_i^2$ and $h_{ijk}$ of
\eqref{ssbterms} is $m_i^2=1\,\,\text{and}\,\,h_{ijk}=\epsilon_{ijk}$.
Therefore, \thee scalar potential, \eqref{totalscalarpotential},
takes \thee form:
\begin{equation}
  V=\frac{1}{4}(F^{ij})^\dag F^{ij}+V_D\,,\label{twistedpotential}
\end{equation}
where $F^{ij}$ is defined as:
\begin{equation}
  F^{ij}=[\phi^i,\phi^j]-i\epsilon^{ijk}(\phi^k)^\dag\,.\label{fuzzyfieldstrength}
\end{equation}
The first term of, \eqref{twistedpotential}, is positive, therefore, \thee global minimum of \thee potential is:
\begin{equation}
  [\phi^i,\phi^j]=i\epsilon_{ijk}(\phi^k)^\dag\,,\quad
  \phi^i(\phi^j)^\dag=R^2\,,\label{twistedfuzzy}
\end{equation}
where $(\phi^i)^\dag$ denotes \thee hermitian conjugate of $\phi^i$ and $[R^2,\phi^i]=0$. It is clear that \thee above relations are related to a fuzzy sphere. This gets even more transparent, after the  consideration of \thee untwisted fields, $\tilde{\phi}^i$, defined \by:
\begin{equation}
  \phi^i=\Omega\tilde{\phi}^i\,,\label{twisted-untwisted}
\end{equation}
where $\Omega\neq1$ satisfies \thee relations:
\begin{align}
\Omega^3=1\,,\,\,\,\,[\Omega,\phi^i]=0\,,\,\,\,\,\Omega^\dag=\Omega^{-1}\,,\,\,\,\,
(\tilde{\phi}^i)^\dag=\tilde{\phi}^i\,\,\Leftrightarrow\,\,
(\phi^i)^\dag=\Omega\phi^i\,.
\end{align}
Therefore, \eqref{twistedfuzzy} reproduces \thee fuzzy sphere
relations, generated \by $\tilde{\phi}^i$
\begin{equation}
  [\tilde{\phi}^i,\tilde{\phi}^j]=i\epsilon_{ijk}\tilde{\phi}^k\,,\quad\tilde{\phi}^i\tilde{\phi}^i=R^2\,,\label{untwisted}
\end{equation}
demonstrating \thee fact that non-commutative space generated \by
$\phi^i$ is actually a twisted fuzzy sphere, $\tilde{S}_N^2$. Next, configurations of \thee twisted fields, $\phi^i$, can be found, i.e. fields satisfying \eqref{twistedfuzzy}. Such a configuration is:
\begin{equation}
  \phi^i=\Omega(\mathbf{1}_3\otimes\lambda^i_{(N)})\,,
\end{equation}
where $\lambda^i_{(N)}$ are \thee $SU(2)$ generators in \thee
$N$-dimensional irreducible representation and $\Omega$ is \thee
matrix:
\begin{equation}
  \Omega=\Omega_3\otimes\mathbf{1}_N\,,\quad\Omega_3=\left(
                                                       \begin{array}{ccc}
                                                         0 & 1 & 0 \\
                                                         0 & 0 & 1 \\
                                                         1 & 0 & 0 \\
                                                       \end{array}
                                                     \right)\,,\quad\Omega^3=\mathbf{1}\,.
\end{equation}
According to \thee transformation \eqref{twisted-untwisted}, \thee
``off-diagonal'' orbifold sectors, \eqref{repaftrprojection}, take \thee block-diagonal form:
\begin{equation}
  \phi^i=\left(
           \begin{array}{ccc}
             0 & (\lambda_{(N)}^i)_{(N,\bar{N},1)} & 0 \\
             0 & 0 & (\lambda_{(N)}^i)_{(1,N,\bar{N})} \\
             (\lambda_{(N)}^i)_{(\bar{N},1,N)} & 0 & 0 \\
           \end{array}
         \right)=\Omega\left(
                         \begin{array}{ccc}
                           \lambda^i_{(N)} & 0 & 0 \\
                           0 & \lambda^i_{(N)} & 0 \\
                           0 & 0 & \lambda^i_{(N)} \\
                         \end{array}
                       \right)\,.\label{triaseena}
\end{equation}
The untwisted fields generating \thee ordinary fuzzy sphere, $\tilde{\phi}^i$, are in block-diagonal form. Each block is considered as a fuzzy sphere, since each one satisfies \thee corresponding commutation relations \eqref{untwisted}. In turn, \thee above configuration in \eqref{triaseena}, that is \thee vacuum of \thee theory, has \thee form of three fuzzy spheres, with relative angles $2\pi/3$. Concluding, \thee solution $\phi^i$ can be viewed as \thee twisted equivalent of three fuzzy spheres, in accordance to \thee orbifolding. Note that \thee $F^{ij}$ of \eqref{fuzzyfieldstrength}, can be interpreted as \thee field strength tensor of \thee spontaneously generated fuzzy extra dimensions. The term $V_D$ of \thee potential induces a change on \thee radius of \thee sphere (in a similar way to \thee case of \thee ordinary fuzzy sphere \cite{aschieri-grammatikopoulos, chatzistavrakidis-steinacker-zoupanos, steinacker}).

\subsection{Chiral models after \thee fuzzy orbifold projection - The $SU(3)_c\times SU(3)_L\times SU(3)_R$ model}

The resulting groups after \thee orbifold projection are various because of \thee different ways \thee gauge group $SU(3N)$ is spontaneously broken. The minimal, anomaly free unified models are $SU(4)\times SU(2)\times SU(2)$, $SU(4)^3$ and $SU(3)^3$\footnote{Similar approaches have been studied in \thee framework of YM matrix models \cite{grosse-lizzi}, lacking phenomenological viability.}. Let us focus on \thee breaking of \thee latter,
that is \thee trinification group $SU(3)_c\times SU(3)_L\times
SU(3)_R$ \cite{glashow, rizov} (see also \cite{MaZoup, ma-mondragon-zoupanos, lazarides-panagiotakopoulos,  babu-he-pakvasa, leontaris-rizos} and for a string theory approach see \cite{kim}). At first, \thee integer $N$ has to be decomposed as $N=n+3$. Then, for $SU(N)$, \thee considered embedding is:
\begin{equation}
  SU(N)\supset SU(n)\times SU(3)\times U(1)\,.\label{decomp}
\end{equation}
Therefore, \thee embedding for \thee gauge group $SU(N)^3$ is:
\begin{equation}
  SU(N)^3\supset SU(n)\times SU(3)\times SU(n)\times SU(3)\times
  SU(n)\times SU(3)\times U(1)^3\,.\label{decomposition}
\end{equation}
The three $U(1)$ factors are ignored\footnote{As anomalous gaining
mass \by \thee Green-Schwarz mechanism and therefore they decouple at
\thee low energy sector of \thee theory \cite{lawrence-nekrason-vafa}.} and \thee representations are decomposed according to \eqref{decomposition}, as:
\begin{align}
  &SU(n)\times SU(n)\times SU(n)\times SU(3)\times SU(3)\times
  SU(3)\,,\nonumber\\
 &(n,\bar{n},1;1,1,1)+(1,n,\bar{n};1,1,1)+(\bar{n},1,n;1,1,1)+(1,1,1;3,\bar{3},1)\nonumber\\
 &+(1,1,1;1,3,\bar{3})+(1,1,1;\bar{3},1,3)+(n,1,1;1,\bar{3},1)+(1,n,1;1,1,\bar{3})\nonumber\\
 &+(1,1,n;\bar{3},1,1)+(\bar{n},1,1;1,1,3)+(1,\bar{n},1;3,1,1)+(1,1,\bar{n};1,3,1)\,.
\end{align}
Taking into account \thee decomposition \eqref{decomp}, \thee gauge
group is broken to $SU(3)^3$. Under $SU(3)^3$, \thee
surviving fields transform as:
\begin{align}
  &SU(3)\times SU(3)\times SU(3)\,,\\
  &\left((3,\bar{3},1)+(\bar{3},1,3)+(1,3,\bar{3})\right)\,,
\end{align}
which correspond to \thee desired chiral representations of \thee trinification group. Under $SU(3)_c\times SU(3)_L\times SU(3)_R$,
\thee quarks and leptons of \thee first family transform as:
\begin{align}
  q=\left(
      \begin{array}{ccc}
        d & u & h \\
        d & u & h \\
        d & u & h \\
      \end{array}
    \right)\sim(3,\bar{3},1)\,, q^c=\left(
                                             \begin{array}{ccc}
                                               d^c & d^c & d^c \\
                                               u^c & u^c & u^c \\
                                               h^c & h^c & h^c \\
                                             \end{array}
                                           \right)\sim(\bar{3},1,3)\,,
                                           \lambda=\left(
                                                     \begin{array}{ccc}
                                                       N & E^c & \text{v} \\
                                                       E & N^c & e \\
                                                       \text{v}^c & e^c & S \\
                                                     \end{array}
                                                   \right)\sim(1,3,\bar{3})\,,
\end{align}
respectively. Matrices for the other two families come in a similar way. It is worth noting that this theory can be upgraded to a two-loop finite theory (for reviews see \cite{MaZoup, inspire2, inspire4,
heinemeyer18}) and give testable predictions \cite{MaZoup}, too. Additionally, fuzzy orbifolds can be used to break spontaneously \thee unification gauge group down to MSSM and then to \thee $SU(3)_c\times U(1)_{em}$. Summarizing this section, we conclude that fuzzy extra dimensions can be used for constructing chiral, renormalizable and phenomenologically viable field-theoretical models.

A natural extension of \thee above ideas \andd methods have been
reported in ref \cite{chatzi-stein-zoup} (see also
\cite{chatzi-stein-zoup2}), realized in \thee context of Matrix
Models (MM). At a fundamental level, \thee MMs introduced \by
Banks-Fischler-Shenker-Susskind (BFSS) \andd
Ishibashi-Kawai-Kitazawa-Tsuchiya (IKKT), are supposed to provide a
non-perturbative definition of M-theory \andd type IIB string theory
respectively \cite{ishibasi-kawai, banks}. On \thee other hand, MMs
are also useful laboratories for \thee study of structures which
could be relevant from a low-energy point of view. Indeed, they
generate a plethora of interesting solutions, corresponding to
strings, D-branes \andd their interactions \cite{ishibasi-kawai,
chepelev}, \as well \as to non-commutative/fuzzy spaces, such \as
fuzzy tori \andd spheres \cite{iso}. Such backgrounds naturally give
rise to non-abelian gauge theories. Therefore, it appears natural to
pose \thee question whether it is possible to construct
phenomenologically interesting particle physics models in this
framework \as well. In addition, an orbifold MM was proposed \by
Aoki-Iso-Suyama (AIS) in \cite{aoki} \as a particular projection of
\thee IKKT model, \andd it is directly related to \thee construction
described above in which fuzzy extra dimensions arise with
trinification gauge theory \cite{fuzzy}. By $\mathbb{Z}_3$ -
orbifolding, \thee original symmetry of \thee IKKT matrix model with
matrix size $3N\times 3N$ is generally reduced from $SO(9,1)\times
U(3N)$ to $SO(3,1) \times U(N)^3$. This model is chiral \andd has
$D=4$, $\mathcal{N}=1$ supersymmetry of Yang-Mills type \as well \as
an inhomogeneous supersymmetry specific to matrix models. The
$\mathbb{Z}_3$ - invariant fermion fields transform \as
bifundamental representations under \thee unbroken gauge symmetry
exactly \as in \thee constructions described above. In \thee future
we plan to extend further \thee studies initiated in refs
\cite{chatzi-stein-zoup, chatzi-stein-zoup2} in \thee context of
orbifolded IKKT models.\\

Our current interest is to continue in two directions. Given that \thee two approaches discussed here led to \thee $\mathcal{N}=1$ trinification GUT $SU(3)^3$, one plan is to examine \thee phenomenological consequences of these models. The models are different in \thee details but certainly there exists a certain common ground. Among others we plan to determine in both cases \thee spectrum of \thee Dirac and Laplace operators in \thee extra dimensions and use them to study \thee behaviour of \thee various couplings, including \thee contributions of \thee massive Kaluza-Klein modes. These contributions are infinite or finite in number, depending on whether \thee extra dimensions are continuous or fuzzy, respectively. We should note that \thee spectrum of \thee Dirac operator at least in \thee case of $SU(3)/U(1)\times U(1)$ is not known.

Another plan is to start with an abelian theory in ten dimensions and
with a simple reduction to obtain an $\mathcal{N}=(1,1)$ abelian theory in six dimensions. Finally, reducing \thee latter theory over a fuzzy sphere, possibly with Chern-Simons terms, to obtain a non-abelian gauge theory in four dimensions provided with soft supersymmetry breaking terms. Recall that \thee last feature was introduced \by hand in \thee realistic models constructed in \thee fuzzy extra dimensions framework.
\section{Gravity as a gauge theory}
In this section we recall the particularly interesting relation between gravity and gauge theories \cite{Utiyama:1956sy, Kibble:1961ba, MacDowell:1977jt, Kibble:1985sn} with ultimate aim to transfer it to the non-commutative framework. 
\subsection{$4-$dimensional gravity as a gauge theory} 
Employing the vielbein formulation of general relativity, we recall that it can be reproduced -at least at a kinematical level- if considered as a gauge theory of its isometries on the $4-$dimensional Minkowski spacetime, i.e. as a gauge theory of the Poincar\'e algebra, $\mathfrak{iso}(1,3)$, which has ten generators: the four generators of local translations $P_a, a=1,2,3,4,$ and the six Lorentz transformations $M_{ab}$,
satisfying the commutation relations:
\bea
[M_{ab},M_{cd}]=4\eta_{[a[c}M_{d]b]}~,\,\,\,\,\, {[}P_a,M_{bc}]=2\eta_{a[b}P_{c]}~,\,\,\,\,\,
{[}P_a,P_b]=0~,
\eea
where $\eta_{ab}$ is the (mostly plus) Minkowski metric. In order to proceed with the gauging, one has to introduce a gauge field for each generator:
the vielbein $e_{\mu}{}^a$ for translations and the spin connection $\omega_{\mu}{}^{ab}$ for Lorentz transformations. The gauge connection will be:
\begin{equation}
A_{\mu}=e_{\mu}{}^a(x)P_a+\frac{1}{2}\omega_{\mu}{}^{ab}(x)M_{ab}~,
\end{equation}
transforming in the adjoint representation:
\begin{equation}
\d A_{\mu}=\partial_{\mu}\epsilon+[A_{\mu},\epsilon]~,
\end{equation}
where the gauge transformation parameter is:
\begin{equation}
\epsilon=\xi^a(x) P_a+\frac{1}{2} \lambda^{ab}(x)M_{ab}~.
\end{equation}
Therefore, one can calculate the transformations of the gauge fields:
\begin{eqnarray}
\d e_{\mu}{}^a&=&\partial_{\mu}\xi^a+\omega_{\mu}{}^{ab}\xi_b-\lambda^a{}_be_{\mu}{}^b~,
\\
\d \omega_{\mu}{}^{ab}&=&\partial_{\mu}\lambda^{ab}-2\l^{[a}{}_c\omega_{\mu}{}^{cb]}~,
\end{eqnarray}
but also, using the standard formula:
\begin{equation}
R_{\mu\nu}(A)=2\partial_{[\mu}A_{\nu]}+[A_{\mu},A_{\nu}]
\end{equation}
and expanding $R_{\mu\nu}(A)=R_{\mu\nu}{}^a(e)P_a+\frac{1}{2} R_{\m\nu}{}^{ab}(\omega)M_{ab}$, one may result with the curvatures of the gauge fields: 
\begin{eqnarray}
R_{\mu\nu}{}^a(e)&=&2\partial_{[\mu}e_{\nu]}{}^a-2\omega_{[\mu}{}^{ab}e_{\nu]b}~,\\
R_{\mu\nu}{}^{ab}(\omega)&=&2\partial_{[\mu}\omega_{\nu]}{}^{ab}-2\omega_{[\mu}{}^{ac}\omega_{\nu]c}{}^b~.
\end{eqnarray}
The condition for vanishing torsion gives the expression of the spin connection with respect to the vielbein. The dynamics follow from the Einstein-Hilbert action:
\begin{equation}
S_{\text{EH4}}=\frac{1}{2}\int \dd^4x\,\epsilon^{\mu\nu\rho\sigma}\epsilon_{abcd}\, e_{\mu}{}^{a}e_{\nu}{}^b R_{\rho\sigma}{}^{cd}(\omega)~,
\end{equation}
which is not an action that results from gauge theory (Yang-Mills type). Thus, rigorously, only the kinematics of $4-$dimensional gravity is obtained by gauge theory, not its dynamics.

\subsection{$3-$dimensional gravity as a gauge theory}

In the $3-$dimensional case, gravity can be completely described as a gauge theory of the corresponding Poincar\'e algebra, as for the kinematcs as for the dynamics \cite{Witten:1988hc}. The $3-$dimensional Einstein-Hilbert action is: 
\begin{equation} \label{eh3}
S_{\text{EH3}}=\frac{1}{2}\int \dd^3x \,\epsilon^{\mu\nu\rho}\epsilon_{a b c}\, e_{ \mu}{}^{a}
R_{ \nu \rho}{}^{ b c}(\omega)~,
\end{equation}
which, as Witten showed, is identical to a Chern-Simons gauge theory of the Poincar\'e algebra $\mathfrak{iso}(1,2)$. This algebra has six generators, three translations $P_{ a}$ and three Lorentz transformations $M^{ a}=\epsilon^{ a b c}M_{ b c}$, with $ a=1,2,3$. Those generators satisfy the following commutation relations:
\begin{equation}
[M_{ a},M_{ b}]=\epsilon_{ a b c}M^{ c}~,\,\,\,\,\,\,
{[}P_{ a},M_{ b}]= \epsilon_{ a b c}P^{ c}~,\,\,\,\,\,
{[}P_{ a},P_{ b}]=0~.
\end{equation}
Following the same procedure for the gauging as in the $4-$dimensional case, gauge connection and gauge parameters are written as:
\begin{eqnarray}
A_{ \m}=e_{ \mu}{}^{ a}P_{ a}+\omega_{ \m}{}^{ a}M_{ a}~,\,\,\,\,\,
\epsilon=\xi^{ a}P_{ a}+\lambda^{ a}M_{ a}~,
\end{eqnarray}
and then one can calculate the gauge transformations and the curvature of the connection. After the appropriate choice of the quadratic form of the algebra, the resulting Chern-Simons action is identical to the Einstein-Hilbert action, \eqref{eh3}. Furthermore, it was proved that the inclusion of a cosmological constant is also possible but then one has to gauge the dS or AdS algebra in three dimensions, $\mathfrak{so}(3,1)$ and $\mathfrak{so}(2,2)$ respectively. In this case the generators of the translations are not commutative any more, but they satisfy the relation:   
\begin{equation} 
[P_{ a},P_{ b}]=\lambda M_{ a b}~,
\end{equation}
where $\lambda$ is the cosmological constant. 
\subsection{$3-$dimensional gravity as a gauge theory on non-commutative spaces}

Having already studied gauge theories on non-commutative spaces (section 3), and given the strong relation between gravity and gauge theories in three dimensions, our purpose is to study gravity as a gauge theory on non-commutative spaces. In order to accomplish this goal, the first and important step is to identify the non-commutative space because it will be its isometry group the one that one would gauge in order to derive the kinematics and the action. 

A very interesting space is the foliation of the $3-$dimensional Euclidean space by fuzzy spheres, first considered in ref \cite{Hammou:2001cc} (see also \cite{Vitale:2014hca}). Non-commutative coordinates obey the algebra of $SU(2)$, but, unlike the fuzzy sphere case, one does not consider the matrices to be proportional to the $SU(2)$ generators in irreducible representations, but in reducible ones. The consideration of reducible representations results in the construction of large, block-diagonal matrices, with each block being an irreducible representation. Therefore the Hilbert space is:
\begin{equation}
{\mathcal{H}}=\oplus [\ell],\quad \ell=0,1/2,1,\ldots~.
\end{equation}
This fuzzy space is known as $\mathbb{R}^3_\lambda$ and can be viewed as being
given by three operators $X_i$ which satisfy:
\begin{equation}
[X_i,X_j]=i\lambda \epsilon_{ijk}X_k~,
\end{equation}
living in reducible representations of $\mathfrak{su}(2)$ (cf. \cite{Hammou:2001cc}). Allowing $X_i$ to live in a reducible representation is equivalent to considering a sum of fuzzy 2-spheres of different radii. Thus, this can be seen as a discrete foliation of 3D Euclidean space by multiple fuzzy 2-spheres, each being a leaf of the foliation\footnote{In the Lorentzian case there is a similar construction, where the $3-$dimensional spacetime with Lorentzian signature is foliated by fuzzy hyperboloids \cite{Jurman:2013ota}.}. (cf. \cite{DeBellis:2010sy}).

The above space has an $SO(4)$ symmetry \cite{Kovacik:2013yca}, which one would gauge. In this procedure, the need of including additional generators emerges, typical in non-Abelian non-commutative gauge theories, for the anticommutators to close. Therefore, the gauge theory considered in this case is the $U(2)\times U(2)$ in a fixed representation. Then, the procedure followed is the same as in the continuous case, adjusted to the non-commutative framework: First one has to establish the commutation and anticommutation relations of the generators:
\begin{eqnarray}
&& [P_a,P_b]=i\epsilon_{abc}M_c~, \quad [P_a,M_b]=i\epsilon_{abc}P_c~, \quad [M_a,M_b]=i\epsilon_{abc}M_c~,
\\
&& \{P_a,P_b\}=\frac{1}{2} \delta_{ab}\one~,\quad \{P_a,M_b\}=\frac{1}{2} \delta_{ab}\g_5~, \quad \{M_a,M_b\}=\frac{1}{2} \d_{ab}\one~.
\end{eqnarray}
Then, one has to introduce a gauge field for each generator and therefore, the gauge connection is obtained, which modifies the coordinates to their covariant form, that is: 
\begin{equation}
{\cal X}_{\mu}= X_{\mu}\otimes i\one +e_{\mu}{}â\otimes P_a+\omega_{\mu}{}â\otimes M_a+A_{\mu}\otimes i\one+{\widetilde{A}}_{\mu}\otimes \g_5~~,
\end{equation}
The gauge parameter is valued in the Lie algebra, therefore it is defined as:
\begin{equation}
\epsilon=\xi^a\otimes P_a+\l^a\otimes M_a+\epsilon_0\otimes i\one+\widetilde{\epsilon}_0\otimes\g_5~.
\end{equation}
Using the above relations, one may end up with the transformations of the gauge fields. Also, using the covariant coordinates in the following relation:
\begin{equation}
\mathcal{R}_{\mu\nu}=[\mathcal{X}_\mu,\mathcal{X}_\nu]-i\lambda\epsilon_{\mu\nu}^{~~~\rho}\mathcal{X}_\rho\,,
\end{equation}
one obtains the corresponding curvatures. Finally, the action proposed is:
\begin{equation}
S=\text{Tr}i\epsilon^{\mu\nu\rho}\mathcal{X}_\mu\mathcal{R}_{\nu\rho}\,.
\end{equation}
Varying the above action one ends up with the equations of motion.

\section{Conclusions}

Kaluza and Klein gave a very insightful suggestion, in which geometry would lead to $4-$dimensional gauge theories. Due to some difficulties, one was led to start with gauge theories in higher-dimensional theories and using the CSDR scheme (with support from the heterotic string), obtain $4-$dimensional particle models, making contact with phenomenology. Going one step further, the employment of fuzzy coset spaces as extra dimensions gave the virtue of renormalizability to the higher-dimensional theories. Then, in order to find support for these scenarios from another direction, a $4-$dimensional particle model that would incorporate the results of a dimensional reduction was built, in which, after suitable additions, a particular fuzzy space, i.e. the fuzzy sphere, was dynamically generated. Eventually, in order to get closer to a complete picture of all interactions, a $3-$d gravity model was proposed, constructed on non-commutative spaces as a gauge theory of their symmetries. The ultimate aim is to obtain such models in four dimensions, with hopes for better u-v properties.

\begin{acknowledgement}
We acknowledge support by the COST action QSPACE MP1405. G.Z. thanks the MPI for Physics in Munich for hospitality and the A.v.Humboldt Foundation for support.
\end{acknowledgement}


\begin{thebibliography}{99.}%
%
%


%
\bibitem{green-scwarz-witten}
Green M.B., Schwarz J.H., Witten E., Cambridge Monographs on Mathematical Physics,
Cambridge University Press, Cambridge, 1987; Green M.B., Schwarz
J.H., Witten E., Cambridge Monographs on Mathematical
Physics, Cambridge University Press, Cambridge, 1987; Polchinski J.,
Cambridge University Press, Cambridge, 1998; Polchinski J., Cambridge
University Press, Cambridge, 1998; Blumenhagen R., L\"{u}st D.,
Theisen S., Springer, 2013.




\bibitem{gross-harvey}
Gross D.J., Harvey J.A., Martinec E.J., Rohm R., Nuclear Phys. B256 (1985)
253; Gross D.J., Harvey J.A., Martinec E.J., Rohm R., Phys. Rev. Lett. 54 (1985) 502.

\bibitem{forgacs-manton}
Forg\'{a}cs P., Manton N.S., Comm. Math. Phys. 72 (1980) 15-35.

\bibitem{kapetanakis-zoupanos}
Kapetanakis D., Zoupanos G., Phys. Rep. 219 (1992).

\bibitem{kubyshin-mourao}
Kubyshin Yu.A., Mour{\~a}o J.M., Rudolph G., Volobujev I.P.,
Lecture Notes in Physics,
Vol. 349, Springer-Verlag, Berlin, 1989.

\bibitem{scherk-schwarz}
Scherk J., Schwarz J.H., Nuclear Phys. B153 (1979) 61-88.

\bibitem{mantonns}
Manton N.S., Nuclear Phys. B193 (1981) 502-516.


\bibitem{Chapline-Slansky}
Chapline G., Slansky R., Nuclear Phys. B209 (1982) 461-483.


\bibitem{Candelas}
P. Candelas, G. T. Horowitz, A. Strominger, E. Witten, Nucl. Phys. B258, (1985) 46


\bibitem{cardoso-curio}
Cardoso G.L., Curio G., Dall'Agata G., L{\"u}st D., Manousselis P.,
Zoupanos G., Nucl. Phys. B 652 (2003) 5-34, hep-th/0211118;
Strominger A., Nucl. Phys. B274
(1986) 253; L{\"u}st D., Nucl.Phys. B276
(1986) 220; Castellani L., L{\"u}st D.,
Nucl.Phys. B296 (1988) 143.

\bibitem{Irges-Zoupanos}
 K. Becker, M. Becker, K. Dasgupta and P. S. Green, JHEP 0304 (2003) 007, hep-th/0301161;
 K. Becker, M. Becker, P. S. Green, K. Dasgupta and E. Sharpe,
  Nucl. Phys. B678 (2004) 19, hep-th/0310058;
 S. Gurrieri, A. Lukas and A. Micu,
  Phys. Rev. D70 (2004) 126009, hep-th/0408121;
 I. Benmachiche, J. Louis and D. Martinez-Pedrera,
  Class. Quant. Grav. 25 (2008) 135006, arXiv:0802.0410 [hep-th];

 A. Micu,
  Phys. Rev. D 70 (2004) 126002, hep-th/0409008;


 A. R. Frey and M. Lippert,
  Phys. Rev. D 72 (2005) 126001, hep-th/0507202;

 P. Manousselis, N. Prezas and G. Zoupanos,
  Nucl. Phys. B739 (2006) 85, hep-th/0511122;
 A. Chatzistavrakidis, P. Manousselis and G. Zoupanos,
  Fortsch. Phys. 57 (2009) 527, arXiv:0811.2182 [hep-th];

 A. Chatzistavrakidis and G. Zoupanos,
  JHEP 0909 (2009) 077, arXiv:0905.2398 [hep-th];


 B. P. Dolan and R. J. Szabo,
  JHEP 0908 (2009) 038, arXiv:0905.4899 [hep-th];


 O. Lechtenfeld, C. Nolle and A. D. Popov,
  JHEP 1009 (2010) 074, arXiv:1007.0236 [hep-th];

 A. D. Popov and R. J. Szabo,
  JHEP 202 (2012) 033, arXiv:1009.3208 [hep-th];

 M. Klaput, A. Lukas and C. Matti,
  JHEP 1109 (2011) 100, arXiv:1107.3573 [hep-th];


 A. Chatzistavrakidis, O. Lechtenfeld and A. D. Popov,
 JHEP 1204 (2012) 114, arXiv:1202.1278 [hep-th];

 J. Gray, M. Larfors and D. L{\"u}st,
 , JHEP 1208 (2012) 099, arXiv:1205.6208 [hep-th];


 M. Klaput, A. Lukas, C. Matti and E. E. Svanes,
 JHEP 1301 (2013) 015, arXiv:1210.5933
[hep-th];


\bibitem{i-z}
N. Irges and G. Zoupanos, Phys. Lett. B698, (2011) 146, arXiv:hep-ph/1102.2220; N. Irges, G. Orfanidis, G. Zoupanos, arXiv:1205.0753 [hep-ph], PoS CORFU2011 (2011)
105.


\bibitem{Butruille}
Butruille J. -B., arXiv:math.DG/0612655.


\bibitem{manousselis-zoupanos}
P. Manousselis, G. Zoupanos,
Phys.Lett. B518 (2001) 171-180, hep-ph/0106033; P. Manousselis, G.
Zoupanos, Phys.Lett. B504 (2001) 122-130


\bibitem{manousselis-zoupanos2}
P. Manousselis, G. Zoupanos, JHEP 0411
(2004) 025, hep-ph/0406207; P. Manousselis, G. Zoupanos, JHEP 0203 (2002) 002.


\bibitem{connes}
Connes A., Academic Press, Inc., San
Diego, CA, 1994.


\bibitem{madorej}
Madore J., London Mathematical
Society Lecture Note Series, Vol. 257, Cambridge University Press,
Cambridge, 1999.


\bibitem{Madore:1991bw}
  J.~Madore,
  Class.\ Quant.\ Grav.\  {\bf 9} (1992) 69.
  doi:10.1088/0264-9381/9/1/008



\bibitem{buric-grammatikopoulos-madore-zoupanos}
Buric M., Grammatikopoulos T., Madore J., Zoupanos G., JHEP 0604 (2006) 054;
Buric M., Madore J., Zoupanos G., SIGMA 3:125,2007, arXiv:0712.4024
[hep-th].


\bibitem{filk}
T. Filk, Phys.
Lett. B 376 (1996) 53; J. C. V\'{a}rilly and J. M.
Gracia-Bond\'{i}a, Int. J. Mod. Phys. A 14 (1999) 1305
[hep-th/9804001]; M. Chaichian, A. Demichev and P. Presnajder,
Nucl. Phys. B 567 (2000)
360, hepth/ 9812180; S. Minwalla, M. Van Raamsdonk and N. Seiberg,
JHEP 0002 (2000) 020, hep-th/9912072.


\bibitem{grosse-wulkenhaar}
H. Grosse and R. Wulkenhaar, Lett. Math. Phys. 71
(2005) 13, hep-th/0403232.


\bibitem{grosse-steinacker}
H. Grosse and H. Steinacker, Adv. Theor. Math.
Phys. 12 (2008) 605, hep-th/0607235; H. Grosse and H. Steinacker,
Nucl. Phys. B 707 (2005)
145, hep-th/0407089.


\bibitem{connes-lott}
Connes A., Lott J., Nuclear Phys. B Proc. Suppl. 18 (1991), 29-47; Chamseddine A.H.,
Connes A., Comm. Math. Phys. 186
(1997), 731-750, hep-th/9606001; Chamseddine A.H., Connes A.,
Phys. Rev. Lett. 99 (2007),
191601, arXiv:0706.3690.


\bibitem{martin-bondia}
Mart\'{i}n C.P., Gracia-Bond{\'i}a M.J., V{\'a}rilly J.C., Phys. Rep. 294 (1998), 363-406, hep-th/9605001.


\bibitem{dubois-madore-kerner}
Dubois-Violette M., Madore J., Kerner R., Phys. Lett. B217 (1989), 485-488;
Dubois-Violette M., Madore J., Kerner R., Classical Quantum Gravity 6 (1989),
1709-1724; Dubois-Violette M., Kerner R., Madore J.,
J. Math. Phys. 31 (1990), 323-330.


\bibitem{madorejz}
Madore J., Phys. Lett. B 305 (1993),
84-89; Madore J., (Sobotka Castle, 1992), Fund. Theories Phys., Vol.
52, Kluwer Acad. Publ., Dordrecht, 1993, 285-298. hep-ph/9209226.


\bibitem{connes-douglas-schwarz}
Connes A., Douglas M.R., Schwarz A., JHEP (1998), no.2, 003,
hep-th/9711162.


\bibitem{seiberg-witten}
Seiberg N., Witten E., JHEP (1999), no.9, 032, hep-th/9908142.


\bibitem{ishibasi-kawai}
N.Ishibashi, H.Kawai, Y.Kitazawa and A.Tsuchiya, Nucl. Phys. B498 (1997) 467, arXiv:hep-th/9612115.


\bibitem{jurco}
Jur\v{c}o B., Schraml S., Schupp P., Wess J., Eur. Phys. J. C 17 (2000), 521-526,
hep-th/0006246; Jur\v{c}o B., Schupp P., Wess J.,
Nuclear Phys. B 604 (2001), 148-180, hep-th/0102129; Jur{\v c}o B.,
Moller L., Schraml S., Schupp S., Wess J., Eur. Phys. J.
C 21 (2001), 383-388, hep-th/0104153; Barnich G., Brandt F.,
Grigoriev M., JHEP (2002), no.8, 023,
hep-th/0206003.


\bibitem{chaichian}
Chaichian M., Pre{\v s}najder P., Sheikh-Jabbari M.M., Tureanu A.,
Eur. Phys. J. C
29 (2003), 413-432, hep-th/0107055.


\bibitem{camlet}
Calmet X., Jur{\v c}o B., Schupp P., Wess J., Wohlgenannt M., Eur. Phys. J. C 23
(2002), 363-376, hep-ph/0111115; Aschieri P., Jur{\v c}o B., Schupp
P., Wess J., Nuclear Phys. B 651 (2003), 45-70, hep-th/0205214; Behr W.,
Deshpande N.G., Duplancic G., Schupp P., Trampetic J., Wess J.,
Eur.Phys.J.C29: 441-446, 2003.


\bibitem{aschieri-madore-manousselis-zoupanos}
Aschieri P., Madore J., Manousselis P., Zoupanos G., JHEP (2004), no. 4, 034,
hep-th/0310072; Aschieri P., Madore J., Manousselis P., Zoupanos G.,
Fortschr. Phys. 52
(2004), 718-723, hep-th/0401200; Aschieri P., Madore J., Manousselis
P., Zoupanos G., Conference: C04-08-20.1 (2005) 135-146,
hep-th/0503039.


\bibitem{aschieri-grammatikopoulos}
Aschieri P., Grammatikopoulos T., Steinacker H., Zoupanos G.,
JHEP (2006), no. 9, 026,
hep-th/0606021; Aschieri P., Steinacker H., Madore J., Manousselis
P., Zoupanos G., arXiv:0704.2880.


\bibitem{steinacker-zoupanos}
Steinacker H., Zoupanos G., JHEP (2007), no. 9, 017,
arXiv:0706.0398.


\bibitem{chatzistavrakidis-steinacker-zoupanos}
A. Chatzistavrakidis, H. Steinacker and G. Zoupanos, Fortsch.Phys. 58 (2010)
537-552, arXiv:0909.5559 [hep-th].


\bibitem{fuzzy}
A.~Chatzistavrakidis, H.~Steinacker and G.~Zoupanos,
    JHEP 1005 (2010) 100,
  arXiv:hep-th/1002.2606
A.~Chatzistavrakidis and G.~Zoupanos,
     SIGMA 6 (2010) 063, arXiv:hep-th/1008.2049.


\bibitem{Gavriil:2015lka}
  D.~Gavriil, G.~Manolakos, G.~Orfanidis and G.~Zoupanos,
  Fortsch.\ Phys.\  {\bf 63} (2015) 442
  doi:10.1002/prop.201500022
  [arXiv:1504.07276 [hep-th]]; 
  G.~Manolakos and G.~Zoupanos,
  Phys.\ Part.\ Nucl.\ Lett.\  {\bf 14} (2017) no.2,  322.
  doi:10.1134/S1547477117020194; 
  G.~Manolakos and G.~Zoupanos,
  Springer Proc.\ Math.\ Stat.\  {\bf 191} (2016) 203
  doi:10.1007/978-981-10-2636-2-13
  [arXiv:1602.03673 [hep-th]].


\bibitem{Witten:1988hc}
  E.~Witten,
  ``(2+1)-Dimensional Gravity as an Exactly Soluble System,''
  Nucl.\ Phys.\ B {\bf 311} (1988) 46.


\bibitem{ncgrav}
A. Chatzistavrakidis, L. Jonke, D. Jurman, G. Manolakos, P. Manousselis,  G. Zoupanos, ``Noncommutative gauge theory and gravity in three dimensions'', to appear.

\bibitem{Wetterich-Palla}
C. Wetterich, Nucl. Phys. B222, 20 (1983); L. Palla, Z.Phys. C 24, 195 (1984); K. Pilch and A. N.
Schellekens, J. Math. Phys. 25,
3455(1984); P. Forgacs, Z. Horvath and L. Palla, Z. Phys. C30, 261(1986);
K. J. Barnes, P. Forgacs, M. Surridge and G. Zoupanos, Z. Phys. C33, 427(1987).


\bibitem{Chapline-Manton}
G. Chapline and N. S. Manton, Nucl.
Phys. B184, 391(1981); F.A.Bais, K. J. Barnes, P. Forgacs and G.
Zoupanos, Nucl. Phys.
B263, 557(1986); Y. A. Kubyshin, J. M. Mourao, I. P. Volobujev,
Int. J. Mod. Phys. A 4, 151(1989).


\bibitem{Harnad}
J. Harnad, S. Shnider and L. Vinet, J. Math.
Phys. 20, 931(1979); 21, 2719(1980); J. Harnad,
S. Shnider and J. Tafel, Lett. Math. Phys. 4, 107(1980).


\bibitem{Farakos-Koutsoumbas}
K. Farakos, G. Koutsoumbas,M. Surridge and G. Zoupanos, Nucl. Phys. B291,
128(1987); ibid., Phys. Lett. B191, 135(1987).


\bibitem{andrews}
Andrews, R.P. et al. Nucl.Phys. B751 (2006) 304-341 hep-th/0601098 SWAT-06-455


\bibitem{madore-schrami-schup-wess}
Madore J., Schraml S., Schupp P., Wess J., Eur. Phys. J. C 16 (2000) 161-167,
hep-th/0001203.


\bibitem{Harland-kurkcuoglu}
Harland D., Kurk{\c c}uo{\v g}lu S., Nucl. Phys. B 821 (2009), 380-398, arXiv:0905.2338.


\bibitem{kachru-silverstein}
Kachru S., Silverstein E., Phys. Rev. Lett. 80 (1998), 4855-4858, hep-th/9802183.


\bibitem{steinacker1}
Steinacker H., Nuclear Phys. B 679 (2004), 66-98,
hep-th/0307075.


\bibitem{kim}
Kim J.E., Phys. Lett. B 564 (2003), 35-41,
hep-th/0301177; Choi K.S., Kim J.E., Phys. Lett. B 567 (2003), 87-92, hep-ph/0305002.


\bibitem{ArkaniHamed:2001nc}
  N.~Arkani-Hamed, A.~G.~Cohen and H.~Georgi,
  Phys.\ Lett.\ B {\bf 513} (2001) 232
  doi:10.1016/S0370-2693(01)00741-9
  [hep-ph/0105239].
  

\bibitem{ArkaniHamed:2001ca}
  N.~Arkani-Hamed, A.~G.~Cohen and H.~Georgi,
  Phys.\ Rev.\ Lett.\  {\bf 86} (2001) 4757
  doi:10.1103/PhysRevLett.86.4757
  [hep-th/0104005].


\bibitem{maalampi-roos}
J. Maalampi and M. Roos, Phys. Rept. 186 (1990) 53.


\bibitem{brink-schwarz-scherk}
Brink L., Schwarz J.H., Scherk J., Nucl. Phys. B 121 (1977), 77-92; Gliozzi F., Scherk J.,
Olive D.I., Nucl. Phys. B 122 (1977), 253-290.


\bibitem{douglas-greene-morrison}
Douglas M.R., Greene B.R., Morrison D.R., Nuclear Phys. B 506 (1997), 84-106, hep-th/9704151.


\bibitem{bailin-love}
Bailin D., Love A., Phys. Rep. 315 (1999), 285-408.


\bibitem{aldabaz-ibanez}
Aldazabal G., Ib{\'a}{\~n}ez L.E., Quevedo F., Uranga A.M.,
JHEP (2000), no. 8, 002,
hep-th/0005067.


\bibitem{lawrence-nekrason-vafa}
Lawrence A.E., Nekrasov N., Vafa C.,Nuclear Phys. B 533 (1998), 199-209,
hep-th/9803015.


\bibitem{kiritsis}
Kiritsis E., Phys. Rep. 421 (2005), 105-190, Erratum, Phys. Rep. 429
(2006), 121-122, hep-th/0310001.


\bibitem{djouadi}
Djouadi A., Phys. Rep. 459
(2008), 1-241, hep-ph/0503173.


\bibitem{Sohnius:1985qm}
  M.~F.~Sohnius,
  Phys.\ Rept.\  {\bf 128} (1985) 39.
  doi:10.1016/0370-1573(85)90023-7


\bibitem{steinacker}
Steinacker H., Springer
Proceedings in Physics, Vol. 98, Springer, Berlin, 2005, 307-311,
hep-th/0409235.


\bibitem{grosse-lizzi}
H. Grosse, F. Lizzi and H. Steinacker, Phys.Rev. D81 (2010)
085034 , arXiv:1001.2703 [hep-th]; H. Steinacker,
Nucl. Phys. B 810 (2009) 1, arXiv:0806.2032 [hep-th].


\bibitem{glashow}
Glashow S.L., Published in Providence Grand Unif.1984:0088, 88-94.


\bibitem{rizov}
Rizov V.A., Bulg. J. Phys. 8 (1981), 461-477.


\bibitem{MaZoup}
E. Ma, M. Mondragon and G. Zoupanos, JHEP {0412} (2004) 026; S. Heinemeyer, E. Ma, M.
Mondragon and G. Zoupanos, AIP Conf. Proc. 1200 (2010) 568, arXiv:0910.0501 [hep-ph].


\bibitem{ma-mondragon-zoupanos}
Ma E., Mondrag\'{o}n M., Zoupanos G., JHEP (2004), no. 12, 026, hep-ph/0407236.


\bibitem{lazarides-panagiotakopoulos}
Lazarides G., Panagiotakopoulos C., Phys. Lett. B 336 (1994), 190-193, hep-ph/9403317.


\bibitem{babu-he-pakvasa}
Babu K.S., He X.G., Pakvasa S., Phys. Rev.
D 33 (1986), 763-772.


\bibitem{leontaris-rizos}
Leontaris G.K., Rizos J., Phys. Lett. B 632 (2006), 710-716,
hep-ph/0510230.


\bibitem{inspire2}
S. Heinemeyer, M. Mondragon and G. Zoupanos,
Int.J.Mod.Phys. A29 (2014) 18, hep-ph/1430032.


\bibitem{inspire4}
S.~Heinemeyer, M.~Mondragon, N.~Tracas and G.~Zoupanos,
  Springer Proc.\ Math.\ Stat.\  {\bf 111} (2014) 177
  doi:10.1007/978-4-431-55285-7-11
  [arXiv:1403.7384 [hep-ph]].


\bibitem{heinemeyer18}
S. Heinemeyer, M. Mondragon and G. Zoupanos, SIGMA 6 (2010) 049, arXiv:1001.0428
[hep-ph].


\bibitem{chatzi-stein-zoup}
A. Chatzistavrakidis, H. Steinacker, , G. Zoupanos, PoS CORFU2011, PoC:
C11-09-04.1, arXiv:1204.6498 [hep-th].


\bibitem{chatzi-stein-zoup2}
A. Chatzistavrakidis, H. Steinacker, G. Zoupanos, JHEP
1109 (2011) 115, arXiv:1107.0265 [hep-th]


\bibitem{banks}
T. Banks, W. Fischler, S. H. Shenker and L. Susskind, Phys. Rev. D 55 (1997) 5112, hep-th/9610043.


\bibitem{chepelev}
I. Chepelev, Y. Makeenko and K. Zarembo, Phys. Lett. B 400 (1997) 43
[hep-th/9701151]; A. Fayyazuddin and D. J. Smith, Mod. Phys. Lett. A 12 (1997)
1447, hep-th/9701168; H. Aoki, N. Ishibashi, S. Iso, H. Kawai, Y.
Kitazawa and T. Tada, Nucl. Phys. B 565 (2000) 176, hep-th/9908141.


\bibitem{iso}
S. Iso, Y. Kimura, K. Tanaka and K. Wakatsuki, Nucl. Phys. B 604
(2001) 121, hep-th/0101102; Y. Kimura, Prog. Theor. Phys. 106
(2001) 445, hep-th/0103192; Y. Kitazawa, Nucl. Phys. B 642 (2002)
210, hep-th/0207115.


\bibitem{aoki}
H. Aoki, S. Iso and T. Suyama, Nucl. Phys. B 634 (2002) 71, arXiv:hep-th/0203277.


\bibitem{Utiyama:1956sy}
  R.~Utiyama,
  Phys.\ Rev.\  {\bf 101} (1956) 1597.
  doi:10.1103/PhysRev.101.1597

 
\bibitem{Kibble:1961ba}
  T.~W.~B.~Kibble,
  J.\ Math.\ Phys.\  {\bf 2} (1961) 212.
  doi:10.1063/1.1703702


\bibitem{MacDowell:1977jt}
  S.~W.~MacDowell and F.~Mansouri,
  Phys.\ Rev.\ Lett.\  {\bf 38} (1977) 739
   Erratum: [Phys.\ Rev.\ Lett.\  {\bf 38} (1977) 1376].
  doi:10.1103/PhysRevLett.38.1376, 10.1103/PhysRevLett.38.739


\bibitem{Kibble:1985sn}
  T.~W.~B.~Kibble and K.~S.~Stelle,
  In *Ezawa, H. ( Ed.), Kamefuchi, S. ( Ed.): Progress In Quantum Field Theory*, 57-81; and refs therein


\bibitem{Hammou:2001cc}
A.~B.~Hammou, M.~Lagraa and M.~M.~Sheikh-Jabbari,
Phys.\ Rev.\ D {\bf 66} (2002) 025025
doi:10.1103/PhysRevD.66.025025
[hep-th/0110291]. 


\bibitem{Vitale:2014hca}
  P.~Vitale,
  Fortsch.\ Phys.\  {\bf 62} (2014) 825
  doi:10.1002/prop.201400037
  [arXiv:1406.1372 [hep-th]].


 \bibitem{DeBellis:2010sy}
 J.~DeBellis, C.~Saemann and R.~J.~Szabo,
 JHEP {\bf 1104} (2011) 075
 doi:10.1007/JHEP04(2011)075
 [arXiv:1012.2236 [hep-th]].


\bibitem{Kovacik:2013yca}
S.~Kov\'{a}\v{c}ik and P.~Pre\v{s}najder,
J.\ Math.\ Phys.\  {\bf 54} (2013) 102103
doi:10.1063/1.4826355
[arXiv:1309.4592 [math-ph]].


\bibitem{Jurman:2013ota}
D.~Jurman and H.~Steinacker,
JHEP {\bf 1401} (2014) 100
doi:10.1007/JHEP01(2014)100
[arXiv:1309.1598 [hep-th]].

\end{thebibliography}
\end{document}